\documentclass[12pt,oneside, a4paper]{article}

\pdfoutput=1  
 
\ifx\pdfoutput\undefined
\usepackage[dvips,bookmarks=false]{hyperref}	
\else
\usepackage{hyperref}	
\fi
\hypersetup{colorlinks,bookmarksopen,bookmarksnumbered,citecolor=blue,linkcolor=blue,pdfstartview=FitH,urlcolor=blue}

\usepackage{tikz}
\usepackage{amsfonts}

\usepackage{graphicx}
\usepackage{amssymb}
\usepackage{cite}
\usepackage{bm}
\usepackage{amsmath,amsthm}
\usepackage{cleveref}
\usepackage{csquotes}
\usepackage{siunitx}
\usepackage{cancel}

\usepackage{slashed}
\usepackage{multirow}
\usepackage{multicol}
\usepackage{xcolor,colortbl}
\usepackage{booktabs}
\usepackage{subcaption}
\usepackage{tabularx}
\numberwithin{equation}{section}

\usepackage{bbold}
\usepackage{tcolorbox}

\usepackage[toc,page]{appendix}
\definecolor{light-gray}{gray}{0.95}
\usepackage{tcolorbox}
\usepackage{amsfonts}
\usepackage[normalem]{ulem}

\begin{document}
\begin{titlepage}
\begin{center}
        {\Large\bf An $A_4$-Symmetric Double Seesaw for Neutrino Masses and Mixing in Light of JUNO results} 
\vspace{0.5cm}

\renewcommand{\thefootnote}{\fnsymbol{footnote}}
{\bf \hspace*{0.0cm} Swaraj Kumar Nanda}$^a$\footnote[1]{swarajnanda.phy@gmail.com}
{\bf, Maibam Ricky Devi}$^{b}$\footnote[2]{deviricky@gmail.com}
{\bf, Chandini Dash}$^{c}$\footnote[3]{dash25chandini@gmail.com}\\
{\bf R.N. Panda}$^{a}$\footnote[4]{rabinarayanpanda@soa.ac.in}
{\bf, Sudhanwa Patra}$^{d,e}$\footnote[5]{sudhanwa@iitbhilai.ac.in}
\vspace{5mm}
\end{center}
$^a$ {\rm {Department of Physics, ITER, SOA University, Bhubaneswar-751030, India}}\\
$^b$ {\rm {Department of Physics, Gauhati University, Guwahati-781014, India}}\\
$^c$ {\rm {Department of Physics, Utkal University, Vani Vihar, Bhubaneswar-751004, India}}\\
$^d$ {\rm {Department of Physics, Indian Institute of Technology Bhilai, Durg-491002, India}}\\
$^e$ {\rm 
Institute of Physics, Sachivalaya Marg, Bhubaneswar-751005, India}

\vspace{4mm} 
\begin{center}
\abstract{
We investigate the double seesaw mechanism for generating light neutrino masses within a Standard Model extension containing three right-handed neutrinos and three gauge-singlet sterile fermions. The flavour structure is governed by an $A_4$ discrete symmetry, which leads to predictive textures for the Dirac neutrino mass matrix $M_D$, the right-handed--sterile neutrino mixing matrix $M_{RS}$, and the sterile-sector Majorana mass matrix $M_S$. With suitable $A_4$ charge assignments and simple flavon vacuum alignments, the model naturally realizes a leading-order tribimaximal (TBM) mixing pattern corrected by a single complex rotation in the $(1,3)$ sector, resulting in the exact TM$_2$ mixing structure. This framework successfully reproduces the observed reactor mixing angle while providing predictive relations among the neutrino masses, mixing parameters, and CP-violating phases. The recent high-precision JUNO determination of the solar mixing angle, however, places the exact TM$_2$ prediction, $\sin^2\theta_{12}\simeq0.341$, in significant tension with the experimental data. Motivated by this discrepancy, we extend the exact TM$_2$ framework by introducing a small perturbation in the sterile-sector Majorana mass matrix, leading to a generalized TM$_2$ mixing structure. The additional degree of freedom relaxes the rigid TM$_2$ sum rule and allows the solar mixing angle to vary over the experimentally preferred region while preserving the successful predictions for the reactor and atmospheric mixing angles. A comprehensive numerical analysis is performed using the latest JUNO and NuFIT~6.1 neutrino oscillation data. The generalized TM$_2$ framework successfully reproduces the observed neutrino mass spectrum and mixing parameters, predicts characteristic correlations among CP-violating phases and oscillation observables, and yields phenomenologically viable regions for neutrinoless double beta decay. The interplay between the double seesaw mechanism, $A_4$ flavour symmetry, and the generalized TM$_2$ structure therefore provides a predictive and experimentally testable framework for understanding neutrino masses and flavour mixing.
}
\end{center}
\end{titlepage}
\renewcommand{\thefootnote}{\arabic{footnote}}
\setcounter{footnote}{0}
\tableofcontents





\section{Introduction}
The Standard Model (SM) of particle physics has been remarkably successful in describing the electromagnetic, weak, and strong interactions among the elementary particles and almost all of its predictions have been verified to high precision scale in numerous experiments, culminating in the latest discovery of the Higgs boson at the Large Hadron Collider (LHC). Despite the remarkable success of the Standard Model (SM) in describing particle physics up to the electroweak scale, it fails to address many fundamental questions at very theoretical as well as experimental level. The theoretical origin of neutrino masses and mixing remains an outstanding question among one of them. The discovery of neutrino oscillations has firmly established that neutrinos are massive and mix with one another, which is confirmed from various experiments~\cite{SNO:2002tuh, T2K:2019efw, DayaBay:2012fng, DoubleChooz:2011ymz}. These results constitute the first experimental evidence of physics beyond the Standard Model while SM cannot accommodate massive neutrinos in its minimal form. 

The most popular canonical (or Type-I) seesaw mechanism provides a natural way to explain the tiny neutrino masses through the introduction of right-handed neutrinos singlets under SM gauge symmetry~\cite{Minkowski:1977sc, Mohapatra:1979ia, Yanagida:1979as, Gell-Mann:1979vob}. The Type-I seesaw is perhaps the simplest and most well-studied framework for generating Majorana neutrino masses. It extends the Standard Model by adding heavy right-handed neutrino fields $N_{Ri}$ ($i=1,2,3$), which are singlets under the SM gauge group $SU(2)_L \times U(1)_Y$. These fields allow the construction of both Dirac and Majorana mass terms. The Type-I seesaw mechanism thus provides a minimal and elegant explanation of neutrino mass generation, embedding naturally in grand unified theories (GUTs) such as $SO(10)$, where right-handed neutrinos appear automatically in the $16$ representation. Moreover, the heavy Majorana neutrinos introduced in this framework can decay out of equilibrium in the early Universe, generating a lepton asymmetry that is later converted into the observed baryon asymmetry through sphaleron processes, providing a natural realization of leptogenesis~\cite{Fukugita:1986hr}. Alternatively, the type-II~\cite{Magg:1980ut, Schechter:1980gr, Cheng:1980qt, Lazarides:1980nt, Mohapatra:1980yp} and type-III~\cite{Foot:1988aq,He:2012ub} offers an alternative explanation for the origin of neutrino masses by introducing a scalar triplet field $\Delta$ transforming as $(1,3,1)$ (fermion triplet $\Sigma(1,3,0)$) under the SM gauge group. Along with the origin and smallness of neutrino masses, the observed pattern of lepton mixing, mass hierarchy~\cite{SNO:2001kpb,KamLAND:2002uet,Super-Kamiokande:1998kpq} and its nature - whether Dirac \cite{Dirac:1928hu} or Majorana type \cite{Majorana:1937vz}, remain among the most significant questions in neutrino physics. 

The limitation with the canonical seesaw mechanism is that the sub-eV scale of light neutrino masses is linked to a very high scale right-handed neutrinos which is beyond the reach of current or planned experiments. Alternatively, the seesaw scale can be lowered down to TeV range viable for LHC or lowe energy phenomenology within inverse seesaw or linear seesaw variants~\cite{Hirsch:2009mx,Gu:2010xc,Dev:2009aw,Deppisch:2015cua,Humbert:2015yva,Parida:2012sq,Brdar:2018sbk,
ThomasArun:2021rwf,Sahu:2022xkq,Ezzat:2021bzs,Mohapatra:2005wg}. 
In this context, the double seesaw mechanism~\cite{Mohapatra:1986aw,Mohapatra:1986bd} comes with added advantage while resulting an additional layer of suppression through extra singlet fermions, allowing for a lower seesaw scale. In comparison to the usual inverse or linear seesaw mechanism where the light neutrinos are Majorana and heavy neutrinos form pseudo-Dirac pair whereas the double seesaw is attractive in the sense that it naturally accommodates very small Majorana neutrino masses for light neutrinos while allowing large Majorana masses for the heavy states to be at intermediate scales or at TeV scale offering the exciting possibility of testing neutrino mass generation and leptogenesis mechanisms at current or near-future experiments phenomenology~\cite{Malinsky:2005bi,Kersten:2007vk,Patra:2023ltl,Adarsh:2025yar,Patel:2023voj}. 

Incorporating seesaw mechanisms within $A_4$ flavour symmetry has long been a successful approach for explaining the observed neutrino masses and mixing patterns while maintaining predictive Yukawa structures. Type-I seesaw realizations with $A_4$ flavon alignments~\cite{Babu:2002dz,Ma:2004zv,Kang:2005bg} have been shown to generate viable neutrino mixing consistent with oscillation data and to provide characteristic predictions for leptonic CP violation and neutrinoless double beta decay. More recently, modular flavour symmetries, particularly modular $A_4$, have attracted considerable attention since they considerably reduce the number of flavon fields while leading to highly constrained parameter spaces and predictive neutrino phenomenology~\cite{Nanda:2025lem,Devi:2023vpe,RickyDevi:2024ijc,CentellesChulia:2025pyn,Kumar:2023moh,CentellesChulia:2023osj,Nomura:2019xsb,Nomura:2023usj,Nomura:2024nwh,Nomura:2024vus,Nomura:2022mgf,Kashav:2025mch,Nomura:2024atp,Singh:2024imk,Kashav:2022kpk,Kashav:2021zir,Priya:2025wdm,Mohanta:2023tzf,Mishra:2022tnf,Mishra:2022egy,Behera:2020lpd,Behera:2020sfe,Petcov:2024vph,Petcov:2023vws,Petcov:2022fjf,Novichkov:2018yse,Petcov:2018mvm,Petcov:2018snn,Zhang:2024rwv,Abbas:2022slb,Nagao:2021rio,deMedeirosVarzielas:2021pug,Wang:2019xbo,Kobayashi:2019gtp,Zhang:2019ngf,Kobayashi:2019xvz,Nomura:2019lnr,Nomura:2019jxj,Kobayashi:2018wkl,Baur:2024lcc,Chen:2024otk,Ding:2019gof,Ding:2019zxk,Feruglio:2017spp,Kobayashi:2018vbk,Kobayashi:2018scp,Feruglio:2021dte}. In parallel, numerous phenomenological investigations based on conventional $A_4$ flavour symmetry continue to provide viable descriptions of neutrino masses and mixing~\cite{CentellesChulia:2025bcg,Kumar:2025cte,Kumar:2024jot,Devi:2021aaz,Devi:2022scm,Bora:2022jdp,Kashav:2023tmz,Verma:2021koo,Verma:2018lro,Sruthilaya:2019hiu,Sruthilaya:2017mzt,Sahu:2022xkq,Mishra:2020fhy}.

Recently, the high-precision measurement of the solar neutrino oscillation parameters by the JUNO experiment has provided an important benchmark for flavour-symmetry-based models of neutrino masses and mixing~\cite{JUNO:2025gmd,Zhang:2025jnn,Ding:2025dzc,Ding:2025dqd,Petcov:2025aci}. In particular, the improved determination of the solar mixing angle, $\sin^2\theta_{12}$, has placed stringent constraints on the widely studied trimaximal mixing patterns. The exact TM$_2$ mixing scheme, which naturally arises in many flavour models based on non-Abelian discrete symmetries, predicts the well-known sum rule 
$\sin\theta_{12}=1/(\sqrt{3}\cos\theta_{13})$, 
leading to $\sin^2\theta_{12} \simeq 0.341$, which is now in significant tension with the recent JUNO result. This development has motivated a renewed investigation of $A_4$-based flavour models and their possible extensions capable of accommodating the latest oscillation data while preserving their predictive nature.

In this work, we revisit the $A_4$ flavour symmetry in the context of the double seesaw mechanism, which simultaneously provides an explanation for the smallness of neutrino masses, offers viable predictions for neutrinoless double beta decay, and can accommodate the generation of the baryon asymmetry of the Universe through the decay of heavy right-handed neutrinos. We demonstrate that the exact TM$_2$ structure emerges naturally within the $A_4$-symmetric double seesaw framework at leading order. Motivated by the recent JUNO results, we further generalize the TM$_2$ mixing pattern by introducing a small perturbation in the sterile-sector Majorana mass matrix. The resulting generalized TM$_2$ framework relaxes the rigid solar-angle prediction of the exact TM$_2$ scenario and successfully reproduces the current neutrino oscillation data, including the latest JUNO determination of $\sin^2\theta_{12}$.

The predictive nature of non-Abelian discrete flavour symmetries extends beyond model building, since they imply characteristic correlations among neutrino masses, leptonic mixing angles, and CP-violating phases. These predictions can be directly tested in present and future long-baseline neutrino oscillation experiments, thereby providing a powerful experimental probe of flavour-symmetry-based scenarios~\cite{Ballett:2013wya,Ballett:2014dua,Agarwalla:2017wct,Blennow:2020snb,Blennow:2020ncm,Antusch:2007rk}. The present work demonstrates that the recent JUNO results provide a particularly stringent test of the exact TM$_2$ framework and motivate a generalized TM$_2$ realization, which retains the predictive features of the underlying $A_4$ flavour symmetry while remaining fully compatible with current neutrino oscillation data.

Motivated by these developments, we revisit the double seesaw mechanism in an $A_4$-symmetric extension of the Standard Model and investigate its predictions for neutrino masses, flavour mixing, neutrinoless double beta decay, and leptogenesis. Special emphasis is placed on examining the compatibility of the exact TM$_2$ mixing pattern with the latest oscillation data and on constructing a generalized TM$_2$ framework capable of accommodating the recent JUNO determination of the solar mixing angle. We assign appropriate $A_4$ representations to the left-handed lepton doublets, right-handed neutrinos, and sterile fermions, and introduce flavon fields transforming under $A_4$ to construct the Dirac neutrino mass matrix $M_D$, the right-handed--sterile neutrino mixing matrix $M_{RS}$, and the sterile-sector Majorana mass matrix $M_S$ from the $A_4$-invariant Yukawa Lagrangian. The double seesaw mechanism naturally explains the smallness of the light neutrino masses through two successive seesaw suppressions. Integrating out the heavy sterile fermions first generates an effective Majorana mass matrix for the right-handed neutrinos, 
$M_R \simeq -M_{RS} M_S^{-1} M_{RS}^{T}$. 
which subsequently gives rise to the light neutrino mass matrix via the conventional type-I seesaw relation, 
$m_\nu \simeq -M_D M_R^{-1} M_D^{T}$. 
The resulting framework accommodates a broad range of heavy-neutrino mass scales, extending from the GeV scale to the grand-unification scale, and naturally realizes quasi-degenerate right-handed neutrinos suitable for resonant leptogenesis. Furthermore, the underlying $A_4$ flavour symmetry considerably restricts the neutrino mass matrices, leading to predictive correlations among neutrino masses, mixing angles, and CP-violating phases. While the leading-order construction gives rise to the exact TM$_2$ mixing pattern, we demonstrate that a minimal perturbation of the sterile-sector Majorana mass matrix generates a generalized TM$_2$ structure that successfully reproduces the latest JUNO and NuFIT~6.1 oscillation data without compromising the predictive nature of the model.

The paper is organized as follows. In Sec.~\ref{sec:model}, we present the theoretical framework of the $A_4$-symmetric double seesaw model, including the particle content, symmetry assignments, and the construction of the neutrino mass matrices from the $A_4$-invariant Yukawa Lagrangian. In Sec.~\ref{sec:TBM}, we discuss the resulting neutrino mixing structure based on the tribimaximal (TBM) pattern and its generalization through a single complex rotation. The numerical analysis, including the determination of the allowed model parameter space and the constraints from current neutrino oscillation data, is presented in Sec.~\ref{sec:neumerical}. The phenomenological implications of the model, including the predictions for neutrino masses, mixing parameters, neutrinoless double beta decay, and leptogenesis, are discussed in Sec.~\ref{sec:result}. Finally, we summarize our main results and conclude in Sec.~\ref{sec:conclusion}. Additional details, analytical expressions, and supplementary tables are collected in the appendices.

\section{Theoretical Framework}
\label{sec:model}
\subsection{Double Seesaw Mechanism}
The double seesaw mechanism is an elegant extension of the canonical Type-I seesaw that naturally explains the smallness of light neutrino masses through the inclusion of additional gauge-singlet fermions. To implement the double seesaw framework for generating sub-eV light neutrino masses, we extend the minimal lepton sector of the Standard Model by introducing three generations of heavy right-handed neutrinos $N_R$ and three generations of left-handed sterile fermions $S_L$. In the flavor basis $(\nu_L,\,N_R^c,\,S_L)$, the neutral fermion mass matrix takes the generic double seesaw form
\begin{equation}
\mathbb{M}_{\rm DSM}=
\begin{pmatrix}
0 & M_D & 0\\
M_D^T & 0 & M_{RS}\\
0 & M_{RS}^T & M_S
\end{pmatrix}.
\end{equation}

Here, $M_D$ denotes the $3\times3$ Dirac neutrino mass matrix connecting $\nu_L$ and $N_R$, $M_{RS}$ is the mixing mass matrix between the right-handed neutrinos and the sterile fermions, while $M_S$ is the Majorana mass matrix of the sterile neutrinos.

The sterile-sector Majorana mass matrix $M_S$ explicitly breaks lepton number. In the double seesaw framework, it is assumed to constitute the largest mass scale in the neutral fermion sector, satisfying the hierarchy
\begin{equation}
M_S \gg M_{RS} \gg M_D.
\end{equation}

The above hierarchy should be understood as referring to the characteristic mass scales associated with the corresponding matrices (equivalently, their typical singular values or operator norms), rather than to an element-by-element comparison of matrix entries. This well-separated hierarchy ensures the validity of the successive block-diagonalization (two-step seesaw) procedure, in which the sterile neutrinos are first integrated out, followed by the intermediate right-handed neutrinos.

Using the hierarchy $M_S \gg M_{RS} \gg M_D$, the light neutrino masses are generated through two successive seesaw approximations: (i) integrating out the heaviest sterile neutrinos $S_L$, and subsequently (ii) integrating out the intermediate right-handed neutrinos $N_R$. The resulting mass matrices for the light neutrinos, the intermediate right-handed neutrinos, and the sterile neutrinos are given by
\begin{eqnarray}
&&m_\nu =-M_D M_R^{-1}M_D^T=-M_D
\left(
-M_{RS}M_S^{-1}M_{RS}^T
\right)^{-1}
M_D^T,
\\
&&m_N =M_R=-M_{RS}M_S^{-1}M_{RS}^T,
\\
&&m_S = M_S.
\label{eq:doubleseesaw1}
\end{eqnarray}

For the block diagonalization of the neutral fermion mass matrix in the double seesaw framework, we follow the general perturbative block-diagonalization formalism developed by Grimus and Lavoura~\cite{Grimus:2000vj}, which provides a systematic treatment of hierarchical Majorana neutrino mass matrices and allows the effective light and heavy neutrino mass matrices to be obtained order by order in the small expansion parameters.

\subsection{\texorpdfstring{ $A_4$ implementation of Double Seesaw Mechanism}{ A4 implementation of Double Seesaw Mechanism}}
\label{sec:A4}
We briefly discuss the $A_4$ implementation of double seesaw mechanism with minimal extension of SM with three generations of right-handed neutrinos $N_{R_i}$, three sterile neutrinos $S_{L_i}$, scalar flavons $\phi_T,\phi_S, \xi, \xi^{\prime}$ that break $A_4$ symmetry, and the SM Higgs doublet. We also summarize the field transformations of various fields under the SM gauge group $SU(2)_L\times U(1)_Y$ and $A_4$ flavor symmetry used in our DSM construction presented in Table~\ref{tab:fields}. The role of $A_4$ flavor symmetry is two fold: (i) it forbids arbitrary mass matrix structures and thereby constraining the Dirac and Majorana mass terms; (ii) it provides a correlated mass matrices such that the resulting double seesaw generated light neutrino mass matrix is essentially governed by the sterile neutrino mass matrix and a scale factor. 
\begin{table}[htb]
\centering
\begin{tabular}{lcccc}
\toprule
Field & $SU(2)_L$ & $U(1)_Y$ & $A_4$ & Comment \\
\midrule
$\ell_{eL}, \ell_{\mu L}, \ell_{\tau L}$ & 2 & $-1/2$ & $1,1',1''$  & Left-handed leptons \\
$e_{R}$ & 1 & $-1$ & $1,1',1''$  &  Right-handed charged leptons \\
$N_{R}$ & 1 & 0 & $1,1',1''$  &  Right-handed neutrinos \\
$S_{L}$ & 1 & 0 & $3$  & Sterile neutrinos \\
$H$ & 2 & $1/2$ & $1$  &  SM Higgs  \\
$\phi_T,\phi_S,\xi, \xi^\prime$ & 1 & 0 & Flavons ($3$, $3$, $1$, $1^\prime$)  & Break $A_4$ \\
\bottomrule
\end{tabular}
\caption{Representative field assignments under $A_4$ and SM gauge group.}
\label{tab:fields}
\end{table}

\noindent
We choose a minimal $A_4$ transformation in all fields relevant for the implementation of the double seesaw mechanism by assigning single one dimensional $A_4$ representations to the left-handed lepton doublets $\ell_{eL}, \ell_{\mu L}, \ell_{\tau L}$, right-handed charged leptons $e_{R}$, and right-handed neutrinos $N_R$ transforming as ${1}, {1}^{\prime}, {1}^{ \prime \prime}$ as presented in Table \,\ref{tab:fields}. The sterile neutrinos $S_{L}$ transform under this $A_4$ group as a triplet $3$. Here, $A_4$ invariant Lagrangian for implementation of the double seesaw mechanism is given by
\begin{eqnarray}
&&\mathcal{L}_{\rm DSM} =\mathcal{L}_{M_D} + \mathcal{L}_{M_{RS}} + \mathcal{L}_{M_S} \nonumber \\
&\mbox{where,}&\mathcal{L}_{M_{D}} = \alpha_D \overline{\ell_{eL}} \tilde{H} N_{R1} +\beta_D \overline{\ell_{\mu L}}\tilde{H} N_{R2}  + \gamma_D \overline{\ell_{\tau L}} \tilde{H} N_{R3}  + h.c.  \nonumber \\
&&\mathcal{L}_{M_{RS}} = \alpha_{RS} \overline{N_{R1}}  \left(S_L \phi_T\right)_{1} + \beta_{RS} \overline{N_{R2}} \left(S_L \phi_T\right)_{1'} 
+ \gamma_{RS} \overline{N_{R3}}  \left(S_L \phi_T\right)_{1''} +h.c. \nonumber\\
&&\mathcal{L}_{M_S} = \left(\alpha_{S} \phi_S + \beta_S \xi + \gamma_{S} \xi^\prime \right)
\overline{S^c_L}S_L +h.c.
\end{eqnarray}
Additionally, we present, here, Lagrangian term for charged lepton mass matrix as,
\begin{eqnarray}
&&\mathcal{L}_{M_{\ell}} =
   \alpha_\ell \overline{\ell_{e L}} H e_{R}  
   + \beta_\ell \overline{\ell_{\mu L}} H \mu_{R}
  + \gamma_\ell \overline{\ell_{\tau L}} H \tau_{R} + h.c. 
  \end{eqnarray}

\subsubsection{Flavon alignments and structure of mass matrices}
The vacuum alignment of different flavon fields breaking $A_4$ symmetry is very important for flavor structure of various mass matrices involved in double seesaw mechanism (see Appendix~\ref{sec:vacuum-alignment})~\cite{Altarelli:2005yp,Altarelli:2005yx,Ma:2001dn,Grimus:2013tva}. 
To obtain the desired Yukawa structure, the $A_4$ flavour symmetry is supplemented by an auxiliary symmetry, for example a discrete $Z_4\times Z_2$ symmetry (or an equivalent global $U(1)_X$ symmetry), following the standard approach adopted in many $A_4$ flavour models. The auxiliary symmetry assigns different charges to the flavon fields $\phi_T$ and $\phi_S$, thereby allowing $\phi_T$ to generate the Dirac and right-handed--sterile mass terms, while $\phi_S$ contributes only to the sterile Majorana mass matrix. Consequently, unwanted cross-couplings such as $N_RS_L\phi_S$ and $S_LS_L\phi_T$ are forbidden, leading to the predictive Yukawa Lagrangian given below. \\
\noindent
With the vacuum expectation value (VEV) alignments
\[
\langle \phi_T \rangle = v_T(1,0,0), \qquad
\langle \phi_S \rangle = v_S(1,1,1), \qquad
\langle \xi \rangle = u, \qquad
\langle \xi^\prime \rangle = u^\prime,
\]
the $A_4$ flavour symmetry yields diagonal structures for the Dirac neutrino mass matrix $M_D$ and the right-handed--sterile mixing matrix $M_{RS}$, while determining the texture of the sterile Majorana mass matrix $M_S$.
With suitable flavon vacuum expectation values (VEVs) as discussed in the Appendix~\ref{sec:vacuum-alignment}, the following minimal and predictive textures of mass matrices are constructed as follows:
\begin{itemize}
\item \textbf{Dirac neutrino mass matrix $M_{D}$:} Using $A_4$ assignment of left-handed lepton doublets $\{l_{e_L}, l_{\mu_L}, l_{\tau_L} \}$ and right-handed neutrinos $\{N_{R1}, N_{R2}, N_{R3}\}$ as ${1}, {1}^{\prime}, {1}^{ \prime \prime}$ as presented in Table.\ref{tab:fields}, the derived structure of Dirac neutrino mass matrix $M_D$ is diagonal as 
\begin{equation}
M_{D} = v\,\mbox{diag}(\alpha_D,\beta_D,\gamma_D)
\end{equation}
where, $\langle H \rangle =v$ is the electroweak VEV. With the above assignments, the Dirac mass matrix is diagonal at leading order and may even reduce to an identity matrix if $\alpha_D=\beta_D=\gamma_D \equiv \kappa_D$. This results from the trivial contraction of $A_4$ singlets without the need for additional flavons. 
\item \textbf{Right–sterile mixing mass matrix $M_{RS}$:} 
The mixing term between the right-handed neutrinos and the sterile neutrinos ($M_{RS}$) is forbidden at tree level by appropriate choice of $A_4$ assignment. The coupling of right-handed neutrinos to sterile fermions proceeds only through flavon insertions. After vacuum alignment, $M_{RS}$ is also diagonal, and in our present analysis, proportional to $M_D$. This alignment is central to the ``screening mechanism'' to be used for double seesaw mechanism which will be discussed in the later section of neutrino masses and mixing. The diagonal structure of $M_{RS} $ after suitable VEV alignment as,
\begin{equation}
M_{RS} = v_T\,\mbox{diag}(\alpha_{RS},\beta_{RS},\gamma_{RS}),
\end{equation}
Similarly, the mass matrix $M_{RS}$ is considered to be diagonal at leading order and may even reduce to an identity matrix if $\alpha_{RS}=\beta_{RS}=\gamma_{RS} \equiv \kappa_{RS}$. 
\item \textbf{Sterile Majorana mass matrix $M_{S}$:} 
The sterile neutrinos $\{S_{L1}, S_{L2}, S_{L3}\}$ are transforming as $A_4$ triplet and the sterile Majorana mass matrix is flavon-dominated and admits the typical structure diagonalized by $U_{\rm TBM}$ with small $U_{13}$ corrections. This is mainly because of the ratio $M_D/M_{RS}^{T}$ is proportional to an identity matrix under screening condition of $M_D$ and $M_{RS}$ assumed in our analysis thereby, the double seesaw mass formula for light neutrinos $m_{\nu} = M_D (M^{-1}_{RS})^T M_S M^{-1}_{RS} M_D^T$ reduces to $m_{\nu} = \kappa_\nu\, M_S = \big(v^2k^2_{D}/v^2_Tk^2_{RS}\big)M_S$.  
Thus, $M_S$ entirely dictates the physical light neutrino spectrum and the resulting structure of $M_S$ is as follows,
\begin{equation}
\label{eq:MS_decomp1}
M_S \;=\; 
\underbrace{\begin{pmatrix}
b+\tfrac{2a}{3} & -\tfrac{a}{3} & -\tfrac{a}{3} \\
-\tfrac{a}{3} & \tfrac{2a}{3} & b-\tfrac{a}{3} \\
-\tfrac{a}{3} & b-\tfrac{a}{3} & \tfrac{2a}{3}
\end{pmatrix}}_{M_S^{(0)}}
\;+\;
\underbrace{\begin{pmatrix}
0 & 0 & d\\
0 & d & 0\\
d & 0 & 0
\end{pmatrix}}_{M_S^{\rm (corr)}} \,.
\end{equation}
Here, $a=\alpha_S v_S$, 
$b=\beta_S u$ and $d=\gamma_S u^\prime$, where $v_S$, $u$, and $u^\prime$ denote the vacuum expectation values (VEVs) of the flavon fields $\phi_S$, $\xi$, and $\xi^\prime$, respectively. In general, the parameters $a$, $b$, and $d$ are complex. Whenever necessary, we express them in terms of their magnitudes and associated complex phases. It is to be noted that the leading-order term preserves the residual symmetry and is diagonalized by the tribimaximal mixing matrix, while the second term introduces controlled breaking of this residual symmetry, generating a nonzero reactor angle and leptonic CP violation.

\item \textbf{Charged Lepton mass matrix $M_{\ell}$:} The diagonal structure of $M_{\ell}$, from the trivial contraction of $A_4$ singlets without the need for additional flavons, is given by
\begin{equation}
M_{\ell} = v \;\mbox{diag}(\alpha_\ell,\beta_\ell,\gamma_\ell)\,.
\end{equation}
The values of $(\alpha_\ell,\beta_\ell,\gamma_\ell)$ are decided by the usual charged lepton Yukawa couplings $(y_e,y_\mu,y_\tau)$. Thus, the structure of charged lepton mass matrix is already diagonal predicting $U_\ell = \mathbf{I}$. This simplifies the relation $U_{\rm PMNS} = U^\dagger_\ell U_\nu \equiv U_\nu$ and thus, the $A_4$ driven predictive mixing matrix for light neutrinos can be compared with the experimentally known $U_{\rm PMNS}$ providing important correlations between neutrino observables and input model parameters in a minimal and predictive way. 
\end{itemize}

We next intend to discuss how the entire flavor structure of $m_\nu$ is governed solely by the sterile-sector mass matrix $M_S$, while the overall neutrino mass scale is controlled by the ratio of the Dirac and intermediate mass scales. This provides a natural and symmetry-driven explanation for the smallness and simplicity of the light neutrino mass matrix in the double seesaw framework. Following Refs.~\cite{Smirnov:1993af,Altarelli:2004za,Lindner:2005pk,Ludl:2015tha,Bajc:2016eiw,Smirnov:2018luj}, we focus in the present work on a particularly well-motivated and predictive realization in which $M_D$ and $M_{RS}$ share the same flavor structure. The screening mechanism naturally emerges in the double seesaw framework when the Dirac mass matrix $M_D$ and the right-handed--sterile mixing matrix $M_{RS}$ possess the same flavor structure. The general double seesaw formula is given by
\begin{eqnarray}
m_\nu
=
M_D(M_{RS}^{-1})^TM_SM_{RS}^{-1}M_D^T .
\label{eq:dss_general}
\end{eqnarray}
Therefore, the relevant quantity entering the screening relation is
$M_D(M_{RS}^{-1})^T$, rather than simply
$M_DM_{RS}^{-1}$. 
The screening condition requires that $M_D$ and $M_{RS}$ be proportional, $M_{RS}=\kappa\,M_D$, 
where $\kappa$ is a real proportionality constant. In general, $(M_{RS}^{-1})^T
=
\kappa^{-1}(M_D^{-1})^T$.

In the present work, motivated by the underlying $A_4$ flavor symmetry and the vacuum alignments discussed in Appendix~C, we consider the minimal realization in which both matrices are diagonal and, in the degenerate Yukawa limit, proportional to the identity,
\begin{eqnarray}
M_D &=& v\,\kappa_D\,\mathbf{I}, \nonumber\\
M_{RS} &=& v_T\,\kappa_{RS}\,\mathbf{I},
\end{eqnarray}
with $\kappa_D$ and $\kappa_{RS}$ being real dimensionless constants. Since both matrices are diagonal, $(M_{RS}^{-1})^T=M_{RS}^{-1}$, 
and consequently $M_D(M_{RS}^{-1})^T
=(v\kappa_D)/(v_T\kappa_{RS})\,
\mathbf{I}$. 
Substituting this relation into Eq.~(\ref{eq:dss_general}) immediately yields
\begin{eqnarray}
m_\nu
=
\left(\frac{v\kappa_D}{v_T\kappa_{RS}}\right)^2
M_S
\equiv
\kappa_\nu M_S\, 
\quad 
\kappa_\nu
=
\frac{v^2\kappa_D^2}{v_T^2\kappa_{RS}^2}.
\end{eqnarray}
Thus, all flavor dependence associated with the Dirac and right-handed--sterile sectors cancels, and the complete flavor structure of the light neutrino mass matrix is governed solely by the sterile-sector Majorana mass matrix $M_S$. As a result, once $M_S$ is specified, the mixing angles and CP phases are completely fixed up to an overall mass normalization, rendering the framework highly predictive. 
Although the present analysis adopts diagonal matrices for simplicity and predictivity, the screening mechanism itself is more general and remains valid whenever $M_D$ and $M_{RS}$ are proportional to one another, even if both are non-diagonal. 

This phenomenon, commonly referred to as the \emph{screening mechanism}, provides a natural explanation for both the smallness and the simplicity of the light neutrino mass matrix. Moreover, it enhances the predictivity of the framework, as the low-energy neutrino masses and mixing parameters can be traced directly to the flavor structure of $M_S$. As discussed in Refs.~\cite{Lindner:2005pk,Brdar:2018sbk}, the equality and simultaneous diagonal structure of $M_D$ and $M_{RS}$ may arise as a consequence of residual $Z_2\times Z_2$ symmetries \cite{Ludl:2015tha}. With the inclusion of an additional permutation symmetry acting on the diagonal entries, one can further
justify the degeneracy of their eigenvalues. Such symmetry-based realizations of the screening mechanism have been widely
explored in the literature on double seesaw models
\cite{Patra:2023ltl,Smirnov:1993af,Altarelli:2004za,Lindner:2005pk,Ludl:2015tha,Bajc:2016eiw,Smirnov:2018luj}, and form the theoretical basis for the predictive $A_4$-symmetric framework adopted in this work. 
\section{\texorpdfstring{Mixing Structure from $A_4$: Tribimaximal Form and a Single Complex Rotation}{Mixing Structure from A4: Tribimaximal Form and a Single Complex Rotation}}
\label{sec:TBM}
Using the screening relation
$m_\nu=\kappa_\nu M_S$, 
the flavor structure of the effective light neutrino mass matrix is entirely determined by the sterile-sector Majorana mass matrix $M_S$. Consequently, the neutrino mixing pattern can be obtained directly from the diagonalization of $M_S$. Following the spontaneous breaking of the $A_4$ flavor symmetry, the leading-order contribution to $M_S$ preserves the residual $Z_2$ symmetry and is diagonalized by the tribimaximal (TBM) mixing matrix. The correction proportional to the flavon parameter $d$ softly breaks this residual symmetry, generating non-vanishing $(1,3)$ matrix elements which are responsible for the observed reactor mixing angle and leptonic CP violation. 
For convenience, we introduce the reduced neutrino mass matrix
\begin{equation}
\widehat{m}_\nu
\equiv
\frac{m_\nu}{\kappa_\nu}
=
M_S,
\label{eq:mhat}
\end{equation}
so that all flavor dependence is contained in $\widehat{m}_\nu$. Its explicit structure is
\begin{equation}
\label{eq:MS_decomp}
\widehat{m}_\nu
=
\underbrace{
\begin{pmatrix}
b+\dfrac{2a}{3} & -\dfrac{a}{3} & -\dfrac{a}{3}\\
-\dfrac{a}{3} & \dfrac{2a}{3} & b-\dfrac{a}{3}\\
-\dfrac{a}{3} & b-\dfrac{a}{3} & \dfrac{2a}{3}
\end{pmatrix}
}_{M_S^{(0)}}
+
\underbrace{
\begin{pmatrix}
0 & 0 & d\\
0 & d & 0\\
d & 0 & 0
\end{pmatrix}
}_{M_S^{(\mathrm{corr})}}.
\end{equation}
The first term preserves the residual $Z_2$ symmetry and is diagonalized by the tribimaximal mixing matrix, whereas the second term introduces controlled breaking of this residual symmetry. To derive the mixing structure, we rotate Eq.~(\ref{eq:MS_decomp}) to the TBM basis by defining
\begin{equation}
\widetilde{m}_\nu
\equiv
U_{\rm TBM}^{T}
\widehat{m}_\nu
U_{\rm TBM},
\label{eq:mtilde}
\end{equation}
where
\begin{equation}
U_{\rm TBM}=
\begin{pmatrix}
\sqrt{\dfrac23} & \dfrac1{\sqrt3} & 0\\[2mm]
-\dfrac1{\sqrt6} & \dfrac1{\sqrt3} & -\dfrac1{\sqrt2}\\[2mm]
-\dfrac1{\sqrt6} & \dfrac1{\sqrt3} & \dfrac1{\sqrt2}
\end{pmatrix}.
\end{equation}
A straightforward calculation gives
\begin{equation}
\widetilde{m}_\nu=
\begin{pmatrix}
a+b-\dfrac{d}{2} & 0 &
-\dfrac{\sqrt3}{2}d\\[2mm]
0 & b+d & 0\\[2mm]
-\dfrac{\sqrt3}{2}d & 0 &
a-b+\dfrac{d}{2}
\end{pmatrix}.
\label{eq:MSprime}
\end{equation}
Thus, in the TBM basis only the $(1,3)$ block remains non-diagonal. The complete neutrino mass matrix is therefore diagonalized by a single complex rotation in the $(1,3)$ plane,
\begin{equation}
U_\nu
=
U_{\rm TBM}
U_{13}(\theta,\psi)\,U_m,
\label{eq:Unu}
\end{equation}
where
\begin{equation}
U_{13}(\theta,\psi)=
\begin{pmatrix}
\cos\theta & 0 & e^{-i\psi}\sin\theta\\
0 & 1 & 0\\
-e^{i\psi}\sin\theta & 0 & \cos\theta
\end{pmatrix} \;\; \text{and}\;\; U_m
=
{\rm diag}
\left(1,e^{i\alpha/2},e^{i\beta/2}\right). 
\end{equation}
Here $\alpha$ and $\beta$ are the Majorana phases.
The diagonalization condition is
$U_{13}^{T}\,
\widetilde{m}_\nu\,
U_{13}
=
\mathrm{diag}
\left(
|\widehat m_1|,
|\widehat m_2|,
|\widehat m_3|
\right)$, 
where the eigenvalues are
\begin{align}
|\widehat m_1|
&=
a+\sqrt{b^2+d^2-bd},
\\
|\widehat m_2|
&=
b+d,
\\
|\widehat m_3|
&=
a-\sqrt{b^2+d^2-bd}.
\end{align}
Using the screening relation, the physical light neutrino masses are simply
\begin{equation}
m_i=\kappa_\nu\,\widehat m_i
=\kappa_\nu M_{S_i},
\qquad
(i=1,2,3),
\end{equation}
showing that the entire low-energy neutrino mass spectrum is dictated by the sterile-sector Majorana mass matrix. The normalization factor $\kappa_\nu$ is fixed by requiring consistency with the
experimentally measured atmospheric mass-squared splitting. For normal mass
ordering (NO), this yields
\begin{align}
\kappa_{\nu}^2 &=
\frac{|\Delta m^2_{\mathrm{atm}}|}
     {M_{S_3}^2 - M_{S_1}^2},
\label{eq:kappa_atm}\\
\Delta m^2_{21} &=
\kappa_{\nu}^2\left(M_{S_2}^2 - M_{S_1}^2\right),
\label{eq:solar_prediction}
\end{align}
so that, once $\kappa_\nu$ is determined from the atmospheric scale, the solar
mass-squared difference follows as a prediction of the model. This feature
highlights the predictive power of the screening mechanism: only one mass scale
is fixed by data, while the remaining mass splitting is governed by the
structure of $M_S$.
For convenience, we introduce the dimensionless ratios
\begin{equation}
\lambda_1 e^{i\phi_{db}} \equiv \frac{d}{b}, \qquad
\lambda_2 e^{i\phi_{ab}} \equiv \frac{a}{b}, \qquad b\neq 0,
\end{equation}
so that $a=b\,\lambda_2 e^{i\phi_{ab}}$ and $d=b\,\lambda_1 e^{i\phi_{db}}$. \\
The leptonic mixing matrix is defined as
\begin{equation}
U_{\rm PMNS} = U_\ell^\dagger\, U_\nu,
\label{eq:UMNS-rel}
\end{equation}
where $U_\ell$ and $U_\nu$ diagonalize the charged-lepton and neutrino mass matrices, respectively. Since the charged-lepton mass matrix is diagonal in the present $A_4$
construction, one has $U_\ell = \mathbf{I}$ 
and hence, $U_{\rm PMNS} = U_\nu$. The leptonic mixing matrix therefore inherits the TBM structure at leading order, corrected by a single complex $1$--$3$ rotation and a diagonal Majorana phase matrix. 
Additionally, the PMNS matrix can be expressed in the standard PDG parameterization as
\begin{equation}
U_{\rm PMNS} =
\begin{pmatrix}
c_{12}c_{13} & s_{12}c_{13} & s_{13}e^{-i\delta_{\rm CP}} \\
-c_{23}s_{12}-c_{12}s_{13}s_{23}e^{i\delta_{\rm CP}} &
 c_{12}c_{23}-s_{12}s_{13}s_{23}e^{i\delta_{\rm CP}} &
 s_{23}c_{13} \\
 s_{12}s_{23}-c_{12}c_{23}s_{13}e^{i\delta_{\rm CP}} &
 -c_{12}s_{23}-s_{12}s_{13}c_{23}e^{i\delta_{\rm CP}} &
 c_{23}c_{13}
\end{pmatrix}
U_m
\label{eq:UPMNS_PDG}
\end{equation}
where $(c_\alpha, s_\alpha)\equiv (\cos\theta_\alpha, \sin\theta_\alpha)$. 
It is worth emphasizing that, throughout our discussion of neutrino oscillation observables, we work in the basis where the charged-lepton mass matrix is diagonal. Consequently, the neutrino mixing matrix is identified with the generalized TM$_2$ mixing matrix, $U_\nu \equiv U_{TM_2}$, 
since the additional Majorana phase matrix does not affect neutrino oscillation probabilities. Whenever Majorana-sensitive observables, such as neutrinoless double beta decay, are considered, the complete mixing matrix is understood to be
$U_\nu = U_{{\rm TM}_2}\,U_m$,
where
$U_m=\mathrm{diag}\!\left(1,e^{i\alpha/2},e^{i\beta/2}\right)
$ contains the two physical Majorana phases. 
The resulting mixing matrix is precisely the trimaximal TM$_2$ mixing matrix,
\begin{equation}
U_{{\rm TM}_2}
=
U_{\rm TBM} U_{13}
=
\left(
\begin{array}{ccc}
 \sqrt{\frac23}\cos\theta
 &
 \frac1{\sqrt3}
 &
 \sqrt{\frac23}\,e^{-i\psi}\sin\theta
 \\[2mm]
 -\frac{\cos\theta}{\sqrt6}
 -\frac{e^{i\psi}\sin\theta}{\sqrt2}
 &
 \frac1{\sqrt3}
 &
 \frac{\cos\theta}{\sqrt2}
 -\frac{e^{-i\psi}\sin\theta}{\sqrt6}
 \\[2mm]
 -\frac{\cos\theta}{\sqrt6}
 +\frac{e^{i\psi}\sin\theta}{\sqrt2}
 &
 \frac1{\sqrt3}
 &
 \frac{\cos\theta}{\sqrt2}
 +\frac{e^{-i\psi}\sin\theta}{\sqrt6}
\end{array}
\right).
\label{eq:UTM2}
\end{equation}
Comparing $U_\nu$ with the standard PDG parameterization of the PMNS
matrix immediately yields the correlations among the observable mixing angles
and the model parameters $(\theta,\psi)$. The solar mixing angle follows from
the $(1,2)$ element,
\begin{equation}
\sin\theta_{12}
=
\frac{1}{\sqrt3\,\cos\theta_{13}},
\label{eq:theta12}
\end{equation}
which is the characteristic TM$_2$ sum rule. Likewise, the reactor angle is
determined from the $(1,3)$ element,
\begin{equation}
\sin\theta_{13}
=
\sqrt{\frac23}\,|\sin\theta|,
\label{eq:theta13}
\end{equation}
while the atmospheric mixing angle is obtained from the ratio of the
$(2,3)$ and $(3,3)$ elements,
\begin{equation}
\tan\theta_{23}
=
\frac{
e^{-i\psi}\sin\theta
+
\sqrt3\,\cos\theta
}{
e^{-i\psi}\sin\theta
-
\sqrt3\,\cos\theta
}.
\label{eq:theta23}
\end{equation}
Therefore, all three leptonic mixing angles are completely determined by only
two parameters, namely the rotation angle $\theta$ and the phase $\psi$,
originating from the controlled breaking of the residual $Z_2$ symmetry in
the sterile neutrino sector. 
The Majorana phase matrix is written as 
$U_m={\rm diag} \left(1,e^{i\alpha/2},e^{i\beta/2}\right).$ , 
while the phase $\psi$ associated with the single $(1,3)$ rotation controls the
Dirac CP violation. In general,
\begin{equation}
\delta_{\rm CP}
=
\left\{
\begin{array}{ll}
\psi, & \sin\theta>0,\\[1mm]
\psi\pm\pi, & \sin\theta<0.
\end{array}
\right.
\label{eq:deltacp}
\end{equation}
Using the current NuFIT~6.0 $3\sigma$ interval for the reactor angle,
$\theta_{13}\in[8.18^\circ,\,8.87^\circ]$, Eq.~(\ref{eq:theta13}) implies
\[
\sin\theta\in[0.174,\,0.188],
\]
so that $\sin\theta$ is positive throughout the phenomenologically allowed
region. Consequently,
\begin{equation}
\boxed{\delta_{\rm CP}=\psi.}
\label{eq:deltapsi}
\end{equation}
Furthermore, employing the analytic relations among the parameters of the
sterile-sector mass matrix, one finds
\begin{equation}
\tan\delta_{\rm CP}
=
-\frac{\sin\phi_{db}}
{\lambda_2\cos(\phi_{ab}-\phi_{db})},
\label{eq:tandelta}
\end{equation}
which directly relates the observable Dirac CP phase to the underlying
high-energy phases of the sterile sector. Thus, within the present
double-seesaw realization, the leptonic CP violation is not an independent
parameter but is completely determined by the flavor structure of $M_S$. 

\medskip
\noindent
\textbf{Heavy neutrino spectrum in the double seesaw.}  
Within the $A_4$-symmetric construction considered here, the Dirac and
right-handed--sterile mass matrices are proportional to the identity. As a
result, the effective right-handed neutrino (RHN) mass matrix obtained after
integrating out the sterile states takes the simple form
\begin{equation}
M_R = - M_{RS} M_S^{-1} M_{RS}^T
     = - v_T^2 \kappa_{RS}^2\, M_S^{-1}.
\end{equation}
Since the sterile-sector mass matrix $M_S$ is diagonalized by the same unitary
matrix as the light neutrino mass matrix,
\begin{equation}
U_S \equiv U_\nu = U_{\rm TBM}\,U_{13}\,U_m,
\end{equation}
it follows that the RHN mass matrix is diagonalized by
$U_R = U_\nu^*$. The eigenvalues of $M_R$ are therefore inversely proportional to
those of $M_S$.

Using the relation $M_{S_j} = m_j/\kappa_\nu$, the RHN masses can be written as
\begin{equation}
M_{R_j}
= \frac{v_T^2 \kappa_{RS}^2}{M_{S_j}}
= \frac{v_T^2 \kappa_{RS}^2}{(m_j/\kappa_\nu)}
= \frac{v^2 \kappa_D^2}{m_j},
\qquad (j=1,2,3).
\end{equation}
This inverse scaling between light and heavy neutrino masses is a characteristic
hallmark of the double seesaw mechanism and arises naturally once screening is
realized through the proportionality of $M_D$ and $M_{RS}$.

The hierarchical structure implied by $M_{R_j}\propto 1/m_j$ has important
phenomenological consequences. In particular, the scale and ordering of the RHN
masses are directly correlated with the low-energy neutrino spectrum, providing
a natural setting for thermal leptogenesis~\cite{Adarsh:2025yar,Patel:2024bjn,Patel:2023voj,Davidson:2008bu,Luty:1992un,Covi:1996wh,Dev:2017trv}. The same $A_4$-induced structure
that controls neutrino mixing at low energies thus also governs the properties
of the heavy states responsible for generating the observed matter--antimatter
asymmetry of the Universe. A detailed analysis of leptogenesis within the double seesaw framework
implemented with $A_4$ flavor symmetry is beyond the scope of the present work
and will be presented separately.
\subsection{Derivation of Double Seesaw Neutrino Masses in the Degenerate Yukawa Limit}
\label{subsec:degenerate_yukawa}
In this subsection, we derive the light neutrino mass spectrum and its parametric
structure arising from the double seesaw mechanism in the limit of degenerate
Yukawa couplings.  After integrating out the heavy states, the effective light neutrino mass matrix
$M_\nu$ depends on three complex parameters $a$, $b$, and $d$, which originate
from the $A_4$-symmetric sterile-sector Majorana mass matrix and its controlled
breaking. In the degenerate Yukawa limit, the overall mass scale factorizes and
can be expressed in terms of a single effective parameter,
\begin{equation}
\kappa_\nu \equiv
\left(\frac{v^2}{v_T^2}\right)
\left(\frac{\kappa_D^2}{\kappa_{RS}^2}\right),
\label{eq:kappa_nu_def}
\end{equation}
where $v$ is the Standard Model Higgs vacuum expectation value, $v_T$ denotes
the sterile-sector vacuum expectation value, and $\kappa_D$ and $\kappa_{RS}$
parametrize the strengths of the Dirac and right-handed--sterile couplings,
respectively. Diagonalizing the effective light neutrino mass matrix, the mass eigenvalues are
obtained as
\begin{equation}
\begin{aligned}
m_1 &= \kappa_\nu\left[a-\sqrt{b^2-bd+d^2}\right],\\
m_2 &= \kappa_\nu\,(b+d),\\
m_3 &= \kappa_\nu\left[a+\sqrt{b^2-bd+d^2}\right],
\end{aligned}
\label{eq:mi_degenerate}
\end{equation}
where the square-root structure reflects the mixing between the parameters $b$
and $d$ induced by the off-diagonal structure of the sterile-sector mass matrix.  
To further streamline the analytic structure, we define the simplified 
combinations
\begin{equation}
p \equiv \frac{m_1}{\kappa_\nu}, \qquad
q \equiv \frac{m_2}{\kappa_\nu}, \qquad
r \equiv \frac{m_3}{\kappa_\nu}.
\label{eq:pqr_def}
\end{equation}
In terms of these variables, the relations in
Eq.~(\ref{eq:mi_degenerate}) take the compact form
\begin{equation}
p = a-\sqrt{b^2-bd+d^2}, \qquad
q = b+d, \qquad
r = a+\sqrt{b^2-bd+d^2}.
\end{equation}
These expressions can be inverted to express the underlying model parameters
directly in terms of the physical neutrino masses,
\begin{eqnarray}
&&a = \frac{p+r}{2}\, , \quad 
b = \frac{q}{2}
    +\frac{1}{2}\sqrt{\frac{(r-p)^2}{3}-\frac{q^2}{3}}
\, , \quad
d = \frac{q}{2}
    -\frac{1}{2}\sqrt{\frac{(r-p)^2}{3}-\frac{q^2}{3}}.
\end{eqnarray}
This parametrization is particularly well suited for numerical analyses, as it
establishes a direct and transparent mapping between the experimentally measured
light neutrino masses and the underlying parameters of the $A_4$-symmetric
double seesaw framework.  In the following section, we will perform a comprehensive numerical scan over the underlying model parameters $(a,b,d,\kappa_\nu)$, subject to current experimental constraints on neutrino oscillation observables. By examining the resulting predictions for neutrino masses, mixing angles, and CP-violating phases with global-fit data and the recent high-precision JUNO measurements, we identify the phenomenologically viable regions of the parameter space and extract representative benchmark points.

\subsection{\texorpdfstring{Extended structure of tribimaximal mixing TM$_2$}{Extended structure of tribimaximal mixing TM2}}
\label{sec: beyond TM2}
In the present framework, the structure of the sterile-sector Majorana mass matrix $M_S$, given in Eq.~(\ref{eq:MS_decomp1}), leads to an effective light neutrino mass matrix that is diagonalized by the TM$_2$ mixing pattern. Consequently, the model predicts
\begin{equation}
\sin^2\theta_{12}\simeq 0.341,
\end{equation}
which is in tension with the recent JUNO~\cite{JUNO:2025gmd} determination of the solar mixing angle. \\

This model based prediction of $\sin^2\theta_{12}$ is disfavored at more than $3\sigma$ by the recent JUNO result~\cite{JUNO:2025gmd}:\\
\begin{equation}
\begin{array}{cc}
&\sin^2\theta_{12} = 0.3092 \pm 0.0087       \\ & \Delta m^2_{21}  = (7.50 \pm 0.12)   \times 10^{-5} \; eV^2
\end{array} \;  \Big\} \; \text{at} \;  1\sigma \;  \text{C.I.}
\end{equation}
This tension indicates that the exact TM$_2$ limit arising from the leading-order flavour structure is no longer phenomenologically preferred. It is therefore important to investigate whether small and well-motivated corrections to the sterile-sector mass matrix can generate departures from the exact TM$_2$ prediction while preserving the essential features of the underlying $A_4$-symmetric double seesaw framework. Since the flavour structure of the effective light neutrino mass matrix is entirely governed by the sterile-sector mass matrix through the screening relation ($m_\nu=\kappa_\nu M_S$), it is sufficient to consider small corrections to ($M_S$). We therefore extend the sterile-sector mass matrix as $M_S \rightarrow M_S+\delta M_S$. The correction $\delta M_S$ softly breaks the residual symmetry responsible for the exact TM$_2$ prediction while preserving the essential structure of the underlying $A_4$-symmetric double seesaw framework. We skip the detailed derivation of $\delta M_S$ which can provide required modification to TM$_2$ pattern explaining the JUNO result. Following the Refs.~\cite{Hyodo:2023sku,Fodroci:2025qgq}, the modified structure of TM$_2$ pattern is given by,

\begin{equation}
\widetilde{U}_{{\rm TM}_2} =
\begin{pmatrix}
\sqrt{\dfrac{2}{3}-\varepsilon}\,\cos\theta &
\sqrt{\dfrac{1}{3}+\varepsilon} &
\sqrt{\dfrac{2}{3}-\varepsilon}\,\sin\theta\,e^{-i\psi}
\\[3mm]

-\sqrt{\dfrac{\varepsilon}{2}+\dfrac{1}{6}}\,\cos\theta
-\dfrac{e^{i\psi}\sin\theta}{\sqrt{2}}
&
\sqrt{\dfrac{1}{3}-\dfrac{\varepsilon}{2}}
&
\dfrac{\cos\theta}{\sqrt{2}}
-\sqrt{\dfrac{\varepsilon}{2}+\dfrac{1}{6}}\,
\sin\theta\,e^{-i\psi}
\\[3mm]

\sqrt{\dfrac{\varepsilon}{2}+\dfrac{1}{6}}\,\cos\theta
-\dfrac{e^{i\psi}\sin\theta}{\sqrt{2}}
&
-\sqrt{\dfrac{1}{3}-\dfrac{\varepsilon}{2}}
&
\dfrac{\cos\theta}{\sqrt{2}}
+\sqrt{\dfrac{\varepsilon}{2}+\dfrac{1}{6}}\,
\sin\theta\,e^{-i\psi}
\end{pmatrix}.
\label{eq:modifiedTM2}
\end{equation}
Here, the parameter $\varepsilon$ characterizes the size of the higher-order flavour-symmetry breaking effects and is expected to be suppressed by the ratio of the flavon vacuum expectation value to the cutoff scale of the effective theory. The correction $\delta M_S$ softly breaks the residual symmetry responsible for the exact TM$_2$ prediction while preserving the essential structure of the underlying $A_4$-symmetric double seesaw framework. The generalized TM$_2$ mixing matrix introduced above leads to modified correlations among the neutrino mixing angles and CP phase. Comparing $U_\nu$ where the modified $TM_2$ is now incorporated, with the standard PMNS parametrization, one obtains

\begin{eqnarray}
&&\sin^2\theta_{13} =
\left(\frac{2}{3}-\varepsilon\right)\sin^2\theta 
\,, \quad 
\sin^2\theta_{12} =
\frac{\frac{1}{3}+\varepsilon}
{1-\left(\frac{2}{3}-\varepsilon\right)\sin^2\theta},
\label{eq:s12_modified} \\
&&\sin^2\theta_{23} = 
\frac{
-\sqrt{3(3\varepsilon+1)}
\sin(2\theta)\cos\psi
+
(3\varepsilon+1)\sin^2\theta
+
3\cos^2\theta
}
{
6-2(2-3\varepsilon)\sin^2\theta
},
\label{eq:s23_modified} \\
&&
\tan\delta = 
\frac{
4+3\varepsilon+(2-3\varepsilon)\cos2\theta
}
{
2-3\varepsilon+(4+3\varepsilon)\cos2\theta
}
\tan\psi.
\label{eq:delta_modified}
\end{eqnarray}
It is important to note that, in the limit $\varepsilon \rightarrow 0$, Eqs.~(\ref{eq:s12_modified})--(\ref{eq:delta_modified}) reduce to the standard TM$_2$ relations obtained in the original framework. In particular, Eq.~(\ref{eq:s12_modified}) reproduces the well-known TM$_2$ sum rule,
$\sin\theta_{12}=1/(\sqrt{3}\cos\theta_{13})$, 
which predicts the narrow range $\sin^2\theta_{12}\simeq0.341$ for the experimentally allowed values of $\theta_{13}$ which is in tension with the recent high-precision JUNO measurement, which favors a significantly smaller value of the solar mixing angle. The introduction of the perturbation parameter $\varepsilon$ generalizes the TM$_2$ mixing structure by relaxing this rigid correlation, thereby allowing $\sin^2\theta_{12}$ to vary within the experimentally preferred region. Consequently, the generalized TM$_2$ framework successfully reconciles the model with the latest JUNO~\cite{JUNO:2025gmd} and current global neutrino oscillation data while preserving the predictive features of the underlying $A_4$-symmetric double seesaw mechanism.
\\
\\
The generalized TM$_2$ framework exhibits an important phenomenological distinction from the exact TM$_2$ mixing scheme. In the exact TM$_2$ limit ($\varepsilon\rightarrow0$), the atmospheric mixing angle depends only on the two parameters $\theta$ and $\psi$, leading to a rigid prediction for the octant of $\theta_{23}$ once the phenomenologically allowed parameter space is specified. In contrast, the introduction of the perturbation parameter $\varepsilon$ modifies the analytical expression for $\sin^2\theta_{23}$, making it a function of all three parameters $(\varepsilon,\theta,\psi)$. Consequently, the atmospheric mixing angle is no longer constrained by the simple TM$_2$ correlation, and its allowed range is shifted together with the solar sector. While the primary motivation for introducing $\varepsilon$ is to reconcile the model with the recent high-precision JUNO determination of the solar mixing angle, the generalized TM$_2$ structure simultaneously alters the atmospheric-sector predictions. In our numerical analysis, the viable parameter space favours $\sin^2\theta_{23}$ in the range listed in Tables~\ref{tab:NO} and \ref{tab:IO}, demonstrating that the generalized TM$_2$ framework possesses greater phenomenological flexibility than the exact TM$_2$ realization while preserving the predictive nature of the underlying $A_4$-symmetric double seesaw model.
\section{Numerical Analysis and Constraints on Model Parameters}
\label{sec:neumerical}
In this section, we intend to perform a detailed numerical analysis of the neutrino masses and neutrino mixing angles as predicted by the double seesaw mechanism embedded within the $A_4$ flavor-symmetric framework. By performing a comprehensive scan over the fundamental model parameters and imposing the current $3\sigma$ constraints from global neutrino oscillation data, we systematically explore both normal and inverted neutrino mass orderings. This procedure allows us to delineate the phenomenologically viable regions of parameter space, identify representative
benchmark points, and uncover characteristic correlations among neutrino masses, mixing angles, and CP-violating phases that emerge as distinctive signatures of the double seesaw realization of $A_4$ symmetry.
\\
\\
We first investigate the phenomenological predictions of the exact TM$_2$ mixing framework and verify with the latest neutrino oscillation data. Motivated by the tension between the exact TM$_2$ prediction for the solar mixing angle and the recent high-precision JUNO measurement, we then extend the analysis to the generalized TM$_2$ framework by introducing a small perturbation parameter. We demonstrate that this generalized structure successfully accommodates the current neutrino oscillation data while preserving the predictive features of the underlying $A_4$-symmetric double seesaw model. Finally, we present benchmark points of the resulting predictions for the neutrino masses and mixing parameters for normal and inverted ordering of the light neutrinos.

\subsection{\texorpdfstring{Numerical analysis for the exact TM$_2$}{Numerical analysis for the exact TM2}}

The sterile-sector Majorana mass matrix $M_S$, constrained by the underlying $A_4$ flavour symmetry, is parameterized by three independent complex parameters $a$, $b$, and $d$. Their magnitudes are scanned over the ranges:

\begin{equation}
\mbox{NO}: 
\begin{array}{ll}
|a| &\in [4.6\times10^{9},\,3.26\times10^{13}]~\mathrm{eV}, \nonumber\\
|b| &\in [3.04\times10^{9},\,9.40\times10^{12}]~\mathrm{eV}, \nonumber\\
|d| &\in [9.35\times10^{8},\,3.26\times10^{13}]~\mathrm{eV},
\end{array} \; \& \;\; \mbox{IO} : 
\begin{array}{ll}
|a| &\in [2.54\times10^{14},\,4.46\times10^{15}]~\mathrm{eV}, \nonumber\\
|b| &\in [2.486\times10^{14},\,4.38\times10^{15}]~\mathrm{eV}, \nonumber\\
|d| &\in [2.32\times10^{14},\,4.18\times10^{15}]~\mathrm{eV}.
\end{array}
\end{equation}
The phases of these parameters are varied freely. The corresponding parameter ranges are chosen such that the resulting low-energy neutrino observables lie within the $3\sigma$ allowed regions of the NuFIT~6.0 global analysis. Simultaneously, the Yukawa-sector parameters are scanned over the ranges, $\kappa_{RS} \simeq 1.0$ and $\kappa_D\in[1\times10^{-7},\,4\times10^{-7}]$ for both normal and inverted mass orderings. These choices ensure the required mass hierarchy among the matrices, $M_D \ll M_{RS} \ll M_S$, which is essential for the validity of the double seesaw mechanism. Throughout the numerical analysis, the Standard Model Higgs vacuum expectation value is fixed at $v=174~\mathrm{GeV}$, while the flavon vacuum expectation value associated with the generation of the matrix $M_{RS}$ is taken to be $v_T=10~\mathrm{TeV}$. The remaining flavon vacuum expectation values are chosen consistently around this scale in accordance with the underlying $A_4$ flavour symmetry. For each randomly generated parameter set, the effective light neutrino mass matrix $m_\nu$ is constructed using the double seesaw relation and subsequently diagonalized numerically. The resulting light neutrino masses are found to lie within the ranges
\begin{equation}
\mbox{NO}: 
\begin{array}{ll}
m_1 &\in [4.0\times10^{-4},\,3.99\times10^{-3}]~\mathrm{eV}, \nonumber\\
m_2 &\in [8.33\times10^{-3},\,9.82\times10^{-3}]~\mathrm{eV}, \nonumber\\
m_3 &\in [4.96\times10^{-2},\,5.12\times10^{-2}]~\mathrm{eV},
\end{array} \; \Rightarrow \; \begin{array}{ll}
\Delta m^2_{21} \in [6.92,\,8.05]\times10^{-5}~\mathrm{eV}^2, \nonumber\\
\Delta m^2_{31} \in [2.46,\,2.61]\times10^{-3}~\mathrm{eV}^2,
\end{array}
\end{equation}
which correctly reproduced the observed solar and atmospheric mass-squared differences,  and are in excellent agreement with current oscillation data. The leptonic mixing angles, obtained from the numerical diagonalization of the effective light neutrino mass matrix $m_\nu$, are found to lie within the ranges
\begin{equation}
\mbox{NO}:   \begin{array}{ll}
\theta_{12} &\in [35.68^\circ,\,35.75^\circ], \nonumber\\
\theta_{13} &\in [8.18^\circ,\,8.87^\circ], \nonumber\\
\theta_{23} &\in [45.00^\circ,\,51.33^\circ],
\end{array} \; \& \;\; \mbox{IO} : 
\begin{array}{ll} \theta_{12} &\in [35.68^\circ,\,35.76^\circ], \nonumber\\
\theta_{13} &\in [8.24^\circ,\,8.90^\circ], \nonumber\\
\theta_{23} &\in [45.00^\circ,\,51.36^\circ].
\end{array}
\end{equation}
Earlier we have seen that
\begin{eqnarray}
\tan\theta_{23} &\simeq& \frac{e^{-i\psi}\sin\theta+\sqrt{3}\,\cos\theta}{
e^{-i\psi}\sin\theta-\sqrt{3}\,\cos\theta}\,, \quad
\Rightarrow \sin^2\theta_{23} \simeq
\frac12\left(
1+
\frac{\sqrt{3}\,\sin2\theta\,\cos\psi}
{3\cos^2\theta+\sin^2\theta}
\right)\,.
\end{eqnarray}
Since, $3\cos^2\theta+\sin^2\theta =1+2\cos^2\theta>0$, the deviation of $\sin^2\theta_{23}$ from maximal mixing is entirely governed by the numerator factor, $(\sqrt3\,\sin2\theta\,\cos\psi)$. Since both orderings predict $\theta \approx 10^\circ$, thus it eventually gives $\cos\theta > 0$ and  $\sin\theta > 0$. The only remaining factor that the sign of the atmospheric angle now depends on is $\cos \psi $. Our model predicts $\psi \in [-89.99^{\circ},\;  89.99^{\circ}]$. As we know that $\cos(-\psi) =\cos(\psi) =\cos (89.99)=0.00017>0$, hence it is safe to say that $\sin^2 \theta_{23}> 1/2$, placing the atmospheric mixing angle in the higher octant ($\theta_{23} > 45^{\circ}$). Thus, our model can resolve the octant degenaracy problem with a preference for higher octant of $\theta_{23}$.

Although the model predicts the atmospheric mixing angle to lie in the higher octant, the corresponding prediction for the solar mixing angle obtained within the exact TM$_2$ framework is $\sin^2\theta_{12} \simeq 0.341$, which is excluded by the recent JUNO measurement at more than the $3\sigma$ level. This discrepancy motivates us to extend the exact TM$_2$ mixing scheme so that the predicted value of $\sin^2\theta_{12}$ is shifted into the experimentally preferred region while preserving the successful predictions for the remaining oscillation observables. To achieve this, we introduced a generalized TM$_2$ structure in Sec.~3.2 by incorporating a small perturbation to the sterile-sector Majorana mass matrix. As demonstrated below, the resulting numerical analysis shows that the generalized TM$_2$ framework successfully reproduces the current neutrino oscillation data, including the latest JUNO determination of the solar mixing angle.
\subsection{\texorpdfstring{Numerical analysis of the generalized $TM_{2}$ mixing structure}{Numerical analysis of the generalized TM2 mixing structure}}
As discussed in Sec.~3.2, the recent high-precision JUNO measurement of the solar mixing angle motivates a generalization of the exact TM$_2$ mixing structure through the introduction of a real perturbation parameter $\varepsilon$ in the tribimaximal (TBM) mixing matrix. This additional degree of freedom relaxes the rigid TM$_2$ prediction for the solar sector, enabling the model to reproduce both the solar mixing angle and the solar mass-squared difference within the narrow intervals preferred by the latest JUNO (2025) data. Using the generalized TM$_2$ parametrization, we vary the model parameters $k_\nu$, $|a|$, $|b|$, and $|d|$ to obtain the light neutrino mass eigenvalues from Eq.~(3.28). We then perform a numerical scan over the generalized TM$_2$ parameters $\varepsilon$, $\theta$, and $\psi$. For each point in the scanned parameter space, the three leptonic mixing angles are computed using Eq.~(3.35), and only those parameter sets that simultaneously satisfy the latest NuFIT~6.1 oscillation constraints and the recent JUNO measurements of $\sin^2\theta_{12}$ and $\Delta m_{21}^{2}$ are retained. 
A comprehensive summary of the allowed model parameters and neutrino observables within the $3\sigma$ region for normal ordering and inverted ordering are presented in Table~\ref{tab:NO} and  Table \ref{tab:IO} respectively, highlighting the highly constrained nature of the viable parameter space and the strong correlations among neutrino masses, mixing angles, and the underlying model parameters.

\begin{table}[htbp]
\centering
\small
\setlength{\tabcolsep}{7pt}
\renewcommand{\arraystretch}{1.25}
\begin{tabular}{|c|c|c|c|}
\hline
\textbf{Parameter} & \textbf{$3\sigma$ Range} &
\textbf{Parameter} & \textbf{$3\sigma$ Range} \\
\hline
$m_1$ (eV) &
$[0.002,\;0.0026]$ &
$\varepsilon$ &
$[-0.039,\;-0.010]$ \\
\hline
$m_2$ (eV) &
$[0.0088,\;0.0091]$ &
$\psi$ (rad) &
$[0.762,\;1.500]$ \\
\hline
$m_3$ (eV) &
$[0.049,\;0.050]$ &
$\theta$ (rad) &
$[0.172,\;0.190]$ \\
\hline
$\Delta m_{21}^{2}/10^{-5}$ (eV$^{2}$) &
$[7.38,\;7.62]$ &
$\tan\delta$ &
$[0.972,\;14.41]$ \\
\hline
$\Delta m_{32}^{2}/10^{-3}$ (eV$^{2}$) &
$[2.45,\;2.59]$ &
$M_{R_1}$ (GeV) &
$[2.95\times10^{-4},\;2.83]$ \\
\hline
$\kappa_{RS}$ &
$[0.001,\;0.089]$ &
$M_{R_2}$ (GeV) &
$[1.65\times10^{-3},\;15.69]$ \\
\hline
$\kappa_D/10^{-10}$ &
$[0.068,\;7.019]$ &
$M_{R_3}$ (GeV) &
$[6.07\times10^{-3},\;68.35]$ \\
\hline
$|a|$ (GeV) &
$[1.476\times10^{5},\;1.846\times10^{5}]$ &
$M_{S_1}$ (GeV) &
$[1.143\times10^{4},\;1.785\times10^{4}]$ \\
\hline
$|b|$ (GeV) &
$[1.019\times10^{5},\;1.264\times10^{5}]$ &
$M_{S_2}$ (GeV) &
$[5.05\times10^{4},\;6.26\times10^{4}]$ \\
\hline
$|d|$ (GeV) &
$[5.029\times10^{4},\;6.492\times10^{4}]$ &
$M_{S_3}$ (GeV) &
$[2.83\times10^{5},\;3.51\times10^{5}]$ \\
\hline
$\sin^{2}\theta_{12}$ &
$[0.2995,\;0.3303]$ &
$k_{\nu}/10^{-16}$ &
$[1.45,\;1.75]$ \\
\hline
$\sin^{2}\theta_{13}$ &
$[0.0207,\;0.0242]$ &
$M_{RS}$ (GeV) &
$[10.037,\;900]$ \\
\hline
$\sin^{2}\theta_{23}$ &
$[0.4320,\;0.4932]$ &
$M_D$ (eV) &
$[122.142,\;11855]$ \\
\hline
\end{tabular}
\caption{Allowed $3\sigma$ ranges of the model parameters and neutrino oscillation observables for the normal mass ordering (NO) obtained within the double-seesaw $A_4$ flavor framework.}
\label{tab:NO}
\end{table}

\noindent
{\bf Benchmark Point and Matrix Structure:}\,
To illustrate the phenomenological implications of the generalized TM$_2$ framework, we present a representative benchmark point. The benchmark values quoted below are obtained from the numerical analysis performed using the generalized TM$_2$ mixing structure. 

\subsubsection{For Normal Ordering}
For normal ordering, the Dirac and intermediate mass matrices are
\begin{align}
M_D &= v\,\kappa_D\,\mathbf{I}
     \;=\; 3.404\times10^{3}\,\mathbf{I}~\mathrm{eV},\\
M_{RS} &=v_T\,\kappa_{RS}\,\mathbf{I}
     \;=\; 2.58\times10^{11}\,\mathbf{I}~\mathrm{eV}.
\end{align}

The sterile-sector Majorana mass matrix takes the form
\begin{equation}
M_S =
\begin{pmatrix}
2.07 & -0.51 & -1.04 \\
-0.51 & 0.48 & 0.54 \\
-1.04 & 0.54 & 1.02
\end{pmatrix}
\times10^{14}~\mathrm{eV},
\end{equation}
leading to the effective light neutrino mass matrix
\begin{equation}
m_\nu =
\begin{pmatrix}
0.0359 & -0.0088 & -0.018 \\
-0.0088 & 0.0084 & 0.0094 \\
-0.0180 & 0.0094 & 0.0176
\end{pmatrix}
~\mathrm{eV}.
\end{equation}
This matrix is diagonalized by a unitary transformation, yielding eigenvalues and mixing parameters consistent with current experimental constraints. The restricted structure of $m_\nu$ highlights the predictive nature of the double seesaw–invoked $A_4$ framework and sets the stage for the phenomenological discussions presented in the next section. 

One of the salient predictions of the double-seesaw realization of the $A_4$ flavor framework is the emergence of large Majorana masses for the right-handed neutrinos. Such heavy states inherently violate lepton number and can therefore give rise to distinctive experimental signatures, including lepton-number-violating processes, neutrinoless
double-beta decay, and possible collider observables, provided that the right-handed neutrino mass scale lies within the sensitivity reach of current or future experiments. In the present numerical analysis, we explicitly evaluate the right-handed neutrino mass matrix and the corresponding unitary matrix that diagonalizes it, which are given by

\begin{align}
M_R &=
\begin{pmatrix}
6.70\times10^{8} & -1.47\times10^{8}  & 7.63\times10^{8} \\
-1.47\times10^{8} & 3.37\times10^{9} & -1.945\times10^{9} \\
7.63\times10^{8} & -1.945\times10^{9} & 2.46\times10^{9}
\end{pmatrix} \,\mathrm{eV}, \label{eq:MRmatrix}
\\[6pt]
U_R &=
\begin{pmatrix}
 0.827 & -0.40 + 0.389\,i & -0.0026 + 0.0005\,i \\
-0.395 + 0.00004\,i & -0.421 + 0.405\,i & 0.69 - 0.124\,i \\
-0.399 - 0.00004\,i & 0.421 + 0.405\,i & -0.69 + 0.124\,i
\end{pmatrix}.
\label{eq:URmatrix}
\end{align}

Table~\ref{tab:NO} summarizes the allowed ranges of the neutrino oscillation observables together with the underlying model parameters obtained from the numerical scan for the normal mass ordering within the generalized TM$_2$ framework of the double seesaw--invoked $A_4$ model. The quoted intervals correspond to the $3\sigma$ allowed parameter space after imposing the latest NuFIT~6.1 oscillation constraints together with the recent high-precision JUNO measurements of the solar neutrino parameters.

The perturbation parameter $\varepsilon$ introduced in the generalized TM$_2$ structure relaxes the rigid solar-angle prediction of the exact TM$_2$ scenario while preserving the predictive features of the underlying $A_4$ flavour symmetry. Consequently, the model successfully reproduces the observed neutrino mass spectrum and mixing parameters, including the experimentally preferred ranges of $\sin^2\theta_{12}$ and $\Delta m_{21}^{2}$. The resulting light neutrino masses $(m_1,m_2,m_3)$ naturally yield the observed normal mass hierarchy together with mass-squared differences in excellent agreement with the current global oscillation data.

The table further shows that the generalized TM$_2$ framework predicts well-constrained ranges for the underlying model parameters, heavy neutrino masses, and low-energy observables, thereby considerably reducing the viable parameter space compared to the exact TM$_2$ realization. These results demonstrate that the generalized TM$_2$ structure successfully accommodates the latest neutrino oscillation data while retaining the predictive nature of the $A_4$-symmetric double seesaw framework. 
\subsubsection{For Inverted Ordering}
We now turn to study the case of inverted mass ordering (IO). In this scenario, the model parameters entering the sterile-sector Majorana mass matrix $M_S$ are varied within the ranges 
\begin{align}
|a| \;  &  \in [1.45\times10^{5},\,1.80\times10^{5}]~\mathrm{GeV}, \nonumber\\
|b| \; &  \in [1.43\times10^{5},\,1.78\times10^{5}]~\mathrm{GeV}, \nonumber\\
|d| \; &  \in [1.44\times10^{5},\,1.78\times10^{5}]~\mathrm{GeV}.
\end{align}
The remaining parameters associated with the Dirac and intermediate mass matrices are fixed as
$v = 174~\mathrm{GeV}$, $v_T = 10~\mathrm{TeV}$, 
$\kappa_D/10^{-10} \in [0.068,\,7.88]$, and
$\kappa_{RS} \in [0.001,\, 0.089]$.
These choices ensure the validity of the double seesaw hierarchy and compatibility with the NuFIT~6.1 oscillation constraints.

A representative benchmark point illustrating the matrix structures in the inverted ordering case is given by
\begin{align}
M_D &= v\,\kappa_D\,\mathbf{I}
     \;=\; 1.16\times10^{4}\,\,\mathbf{I}~\mathrm{eV}, \\
M_{RS} &=  v_T\,\kappa_{RS}\,\mathbf{I}
     \;=\; 8.99\times10^{11}\,\mathbf{I}~\mathrm{eV}.
\end{align}

The sterile-sector Majorana mass matrix, consistent with the $A_4$ flavor symmetry, takes the form
\begin{equation}
M_S =
\begin{pmatrix}
 2.57 & -0.52 &  1.02 \\
-0.52 &  2.57 &  1.02 \\
 1.02 &  1.02 &  1.038
\end{pmatrix}
\times10^{14}~\mathrm{eV}.
\end{equation}
The effective light neutrino mass matrix is obtained as
\begin{equation}
m_\nu = 
\begin{pmatrix}
 0.043 & -0.0086 &  0.0170 \\
-0.0086 &  0.043 &  0.0170 \\
 0.0170 &  0.0170 &  0.0173
\end{pmatrix}
~\mathrm{eV}.
\end{equation}

Diagonalization of the effective light neutrino mass matrix in the inverted mass ordering resulting in a spectrum with a quasi-degenerate pair $(m_1,m_2)$ and a comparatively lighter third state $m_3$, fully consistent with the $3\sigma$ ranges of the NuFIT~6.1 global analysis. As in the normal ordering case, the inverted hierarchy exhibits strong correlations among neutrino masses, mixing angles, and CP-violating phases, arising from the constrained structure imposed by the double seesaw–invoked $A_4$ symmetry. While these features are implicit in the mass matrices, their phenomenological implications are most transparently revealed through correlation plots. Therefore, we present a detailed graphical analysis in the following section to highlight the interplay between the low-energy neutrino observables and the underlying high-scale model parameters, enabling a clearer assessment of the predictability of the model and its sensitivity to future precision measurements.

\begin{table}[htbp]
\centering
\small
\setlength{\tabcolsep}{7pt}
\renewcommand{\arraystretch}{1.25}
\begin{tabular}{|c|c|c|c|}
\hline
\textbf{Parameter} & \textbf{$3\sigma$ Range} &
\textbf{Parameter} & \textbf{$3\sigma$ Range} \\
\hline
$m_1$ (eV) &
$[0.0485,\;0.0500]$ &
$\varepsilon$ &
$[-0.039,\;-0.010]$ \\
\hline
$m_2$ (eV) &
$[0.0493,\;0.0507]$ &
$\psi$ (rad) &
$[0.8517,\;1.500]$ \\
\hline
$m_3$ (eV) &
$[0.0020,\;0.0025]$ &
$\theta$ (rad) &
$[0.173,\;0.190]$ \\
\hline
$\Delta m_{21}^{2}/10^{-5}$ (eV$^{2}$) &
$[7.2,\;7.8]$ &
$\tan\delta$ &
$[1.163,\;14.39]$ \\
\hline
$\Delta m_{31}^{2}/10^{-3}$ (eV$^{2}$) &
$[-2.56,\;-2.43]$ &
$M_{R_1}$ (GeV) &
$[3.65\times10^{-4},\;2.76]$ \\
\hline
$\kappa_{RS}$ &
$[0.001,\;0.089]$ &
$M_{R_2}$ (GeV) &
$[3.67\times10^{-4},\;2.78]$ \\
\hline
$\kappa_D/10^{-10}$ &
$[0.068,\;7.88]$ &
$M_{R_3}$ (GeV) &
$[0.056,\;729.7]$ \\
\hline
$|a|$ (GeV) &
$[1.45\times10^{5},\;1.80\times10^{5}]$ &
$M_{S_1}$ (GeV) &
$[9.36\times10^{2},\;2.76\times10^{3}]$ \\
\hline
$|b|$ (GeV) &
$[1.43\times10^{5},\;1.78\times10^{5}]$ &
$M_{S_2}$ (GeV) &
$[2.87\times10^{5},\;3.56\times10^{5}]$ \\
\hline
$|d|$ (GeV) &
$[1.44\times10^{5},\;1.78\times10^{5}]$ &
$M_{S_3}$ (GeV) &
$[2.89\times10^{5},\;3.58\times10^{5}]$ \\
\hline
$\sin^{2}\theta_{12}$ &
$[0.2996,\;0.3294]$ &
$k_{\nu}/10^{-16}$ &
$[1.45,\;1.75]$ \\
\hline
$\sin^{2}\theta_{13}$ &
$[0.0209,\;0.0243]$ &
$M_{RS}$ (GeV) &
$[10.94,\;900]$ \\
\hline
$\sin^{2}\theta_{23}$ &
$[0.4370,\;0.4931]$ &
$M_D$ (eV) &
$[137.146,\;11868.4]$ \\
\hline
\end{tabular}
\caption{Allowed $3\sigma$ ranges of the model parameters and neutrino oscillation observables for the inverted mass ordering (IO) obtained within the generalized TM$_2$ framework of the double seesaw--invoked $A_4$ flavor symmetry after including perturbative corrections to the sterile Majorana mass matrix $M_S$.}
\label{tab:IO}
\end{table}

Table~\ref{tab:IO} summarizes the allowed ranges of the neutrino oscillation observables together with the corresponding model parameters obtained from the numerical scan for the inverted mass ordering within the generalized TM$_2$ framework of the double seesaw--invoked $A_4$ model. The quoted intervals correspond to the $3\sigma$ allowed parameter space after imposing the latest NuFIT~6.1 oscillation constraints together with the recent JUNO measurements of the solar neutrino parameters.

For the inverted mass ordering, the model predicts a quasi-degenerate pair of heavier neutrino mass eigenstates, $m_1$ and $m_2$, while the lightest neutrino mass, $m_3$, lies in the sub-meV to few-meV region. The corresponding mass-squared differences, $\Delta m_{21}^{2}$ and $\Delta m_{32}^{2}$, are found to be in excellent agreement with the current global oscillation data. The introduction of the perturbation parameter $\varepsilon$ in the generalized TM$_2$ structure successfully accommodates the experimentally preferred values of $\sin^2\theta_{12}$ and $\Delta m_{21}^{2}$ without affecting the successful description of the remaining oscillation observables.

The table further shows that the underlying model parameters, together with the heavy right-handed neutrino masses $M_{R_i}$ and sterile neutrino masses $M_{S_i}$, retain the characteristic hierarchical structure expected from the double seesaw mechanism. Similar to the normal mass ordering, the generalized TM$_2$ framework provides a highly constrained and predictive parameter space for the inverted ordering, demonstrating that the model remains fully compatible with the latest JUNO and NuFIT~6.1 data while preserving the predictive features of the underlying $A_4$ flavour symmetry. The explicit forms of $M_R$ and $U_R$ for a representative benchmark point are presented below.
\begin{eqnarray}
M_R &=&
\begin{pmatrix}
 8.16\times10^{10} & 7.90\times10^{10} & -1.58\times10^{11} \\
 7.9\times10^{10} & 8.16\times10^{10} & -1.58\times10^{10} \\
 -1.58\times10^{11} & -1.58\times10^{11} & 3.18\times10^{11}
\end{pmatrix}\,\mathrm{eV}, \nonumber\\[6pt]
U_R &=&
\begin{pmatrix}
 0.8337 & 0.475 + 0.28\,i & -0.0016 + 0.002\,i \\
 -0.38 + 0.00005\,i & 0.507 + 0.3006\,i & 0.434 - 0.56\,i \\
 -0.392 - 0.00005\,i & 0.507 + 0.3006\,i & -0.432 + 0.5577\,i
\end{pmatrix}.
\end{eqnarray}
Using the analytical expression given in Eq.~(\ref{eq:s12_modified}), we predict the allowed ranges of the neutrino oscillation mixing angles $\theta_{12}$, $\theta_{12}$  and $\theta_{23}$ within our model. We estimate the value of the three mixing angles by varying the input model parameters, $\theta$ and $\psi$. To examine these predictions with experimental data, we take the $3\sigma$ range of $\theta_{13}$ from the NuFIT~6.1 global analysis~\cite{Esteban:2024eli} as an additional input parameter. As summarized in Tables~2 and~3, the solar mixing angle $\theta_{12}$ is predicted to lie within a remarkably narrow interval, $\sin^2\theta_{12}\in[0.2995, 0.3303]$ for normal ordering and $\sin^2\theta_{12}\in[0.2996, 0.3294]$ for inverted ordering. These model-predicted ranges are well contained within the corresponding NuFIT~6.1 $3\sigma$ allowed region as well as for the recent experimental bounds of JUNO  for both mass orderings.
\section{Results and Discussion}
\label{sec:result}
In this section, we present a detailed phenomenological analysis of neutrino masses and mixing predicted within the double seesaw--invoked $A_4$ flavor symmetry framework. Starting from the effective light neutrino mass matrix derived in the previous section, we extract the neutrino mass eigenvalues, neutrino  mixing angles, and the Dirac CP-violating phase using analytical expressions obtained from the diagonalization procedure. These theoretical predictions are then examined with the most recent global fits of neutrino oscillation data. The free parameters of the model are systematically constrained by imposing the experimentally allowed $3\sigma$ ranges of the neutrino mass-squared differences and mixing angles, as reported by the latest NuFIT global analysis. Both normal ordering (NO) and inverted ordering (IO) of neutrino masses are analyzed separately. The combined effects of the resulting mass structures and the double seesaw mechanism lead to characteristic correlations among neutrino observables, which are explored through extensive numerical scans of the parameter space. The resulting predictions for the absolute neutrino mass scale, the Dirac CP phase, and other low-energy observables are critically assessed by comparing the model predictions with current oscillation data.

For a given set of the underlying model parameters $(a,b,d,\kappa_{RS},\kappa_D)$, the eigenvalues of the light neutrino mass matrix $(m_1,m_2,m_3)$ are computed as summarized in Tables~\ref{tab:NO} and \ref{tab:IO}. These masses determine the neutrino mass-squared differences, which are found to lie within the experimentally allowed ranges. In particular, the model predicts
\begin{equation}
\Delta m_{21}^{2}\in(7.38, \; 7.62)\times10^{-5}\ {\rm eV}^{2},
\end{equation}
for both mass orderings, while the atmospheric mass-squared differences are obtained as
\begin{align}
\Delta m_{31}^{2} \in (2.45,\; 2.59)\times10^{-3}\ {\rm eV}^{2}\,\, {\rm (NO)}\,,
\quad\,
\Delta m_{32}^{2} \in (-2.56, \;-2.43)\times10^{-3}\ {\rm eV}^{2}\,
\, {\rm (IO)}. \nonumber
\end{align}

\begin{figure}[htb!]
\centering
\begin{subfigure}{0.48\textwidth}
\caption{$\phi_{db}$ vs $\psi$}
\includegraphics[width=\linewidth]{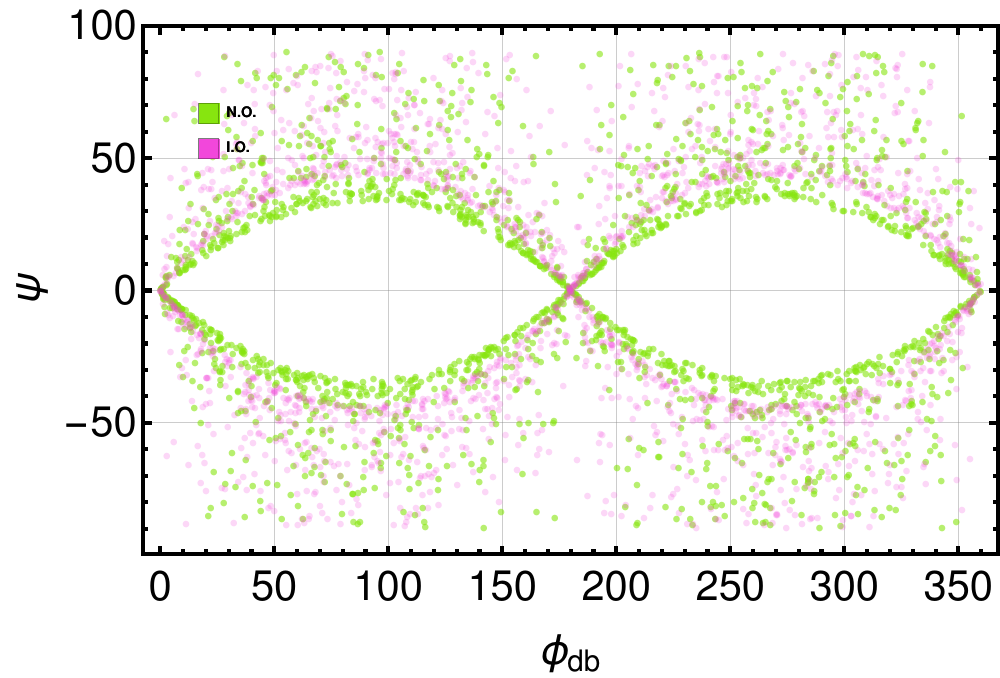}
\end{subfigure}\hfill
\begin{subfigure}{0.48\textwidth}
\caption{$\psi$ vs $\theta_{23}$}
\includegraphics[width=\linewidth]{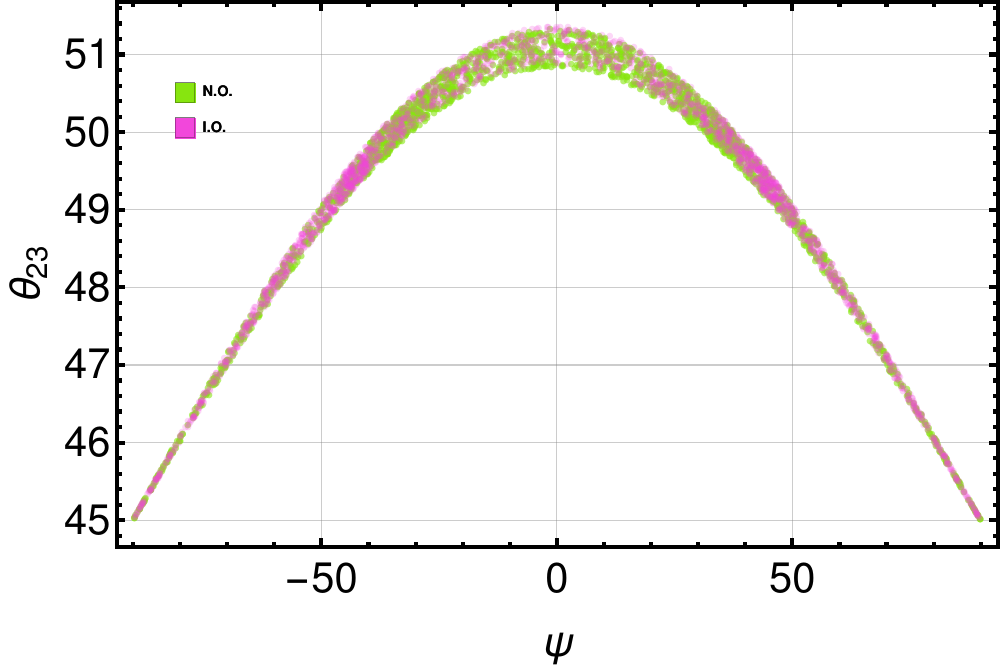}
\end{subfigure}
\caption{Correlation plots among phases and the atmospheric mixing angle in the double seesaw--invoked $A_4$ framework. Panels show (a) $\phi_{db}$ vs $\psi$, and (b) $\psi$ vs $\theta_{23}$. Green (pink) points correspond to normal (inverted) mass ordering.}
\label{fig:3}
\end{figure}
To illustrate the predictive nature of the model, we retain only those correlation plots that reveal non-trivial relationships among the underlying model parameters and measurable neutrino observables. 
Figure~\ref{fig:3} displays the resulting correlations among the relevant model parameters and neutrino oscillation observables obtained from the numerical analysis of exact TM$_2$. Fig.~\ref{fig:3}(a) presents the correlation between the physical phase $\psi$ and the underlying phase $\phi_{db}$ appearing in the sterile-sector Majorana mass matrix. Rather than spanning the entire parameter space, the allowed points are confined to a narrow region, demonstrating that the flavour symmetry significantly restricts the CP-violating phases. Since $\psi$ directly determines the leptonic CP phase in the present framework, this correlation provides a characteristic prediction of the model that can be tested by future precision measurements of leptonic CP violation.
\\
The correlation between the phase parameter $\psi$ and the atmospheric mixing angle is displayed in Fig.~\ref{fig:3}(b). A well-defined band is observed, indicating that the atmospheric mixing angle is not an independent parameter but is strongly correlated with the underlying flavour phase. Moreover, the normal- and inverted-ordering solutions exhibit a substantial overlap, implying that this prediction is largely insensitive to the neutrino mass ordering. Such phase--mixing correlations are a distinctive consequence of the $A_4$-symmetric double seesaw framework and provide an important phenomenological signature that can be scrutinized in future oscillation experiments. Overall, the results presented in this subsection demonstrate that the double seesaw--invoked $A_4$ framework not only reproduces the observed neutrino mass spectrum and mixing angles but also predicts characteristic correlations among CP-violating phases and oscillation observables. These non-trivial correlations considerably enhance the predictive power of the model and motivate the generalized TM$_2$ analysis presented in the following subsection. 
%

\subsection{Impact on Solar Neutrino Parameters from JUNO first data and other experimental solar measurements }
The solar neutrino oscillation parameters provide a stringent test of flavor models, particularly those predicting correlated structures among mixing angles and mass splittings. We intend here to examine whether or not the expected values of the neutrino mixing angles and mass-square differences in our double seesaw-invoked $A_4$ framework are within  the experimental limit set by the ongoing or future planned experiments. 
\\
Recently, the Jiangmen Underground Neutrino Observatory (\textsf{JUNO}~\cite{JUNO:2025gmd}) collaboration reported the first simultaneous high-precision determination of the solar mixing angle $\theta_{12}$ and the mass-squared difference $\Delta m^2_{21}$ assuming normal mass ordering (NO). The reported values are
\begin{equation}
\Delta m^2_{21} = m_2^2 - m_1^2 = (7.50 \pm 0.12)\times 10^{-5}~\text{eV}^2, 
\qquad 
\sin^2\theta_{12} = 0.3092 \pm 0.0087,
\end{equation}
corresponding to sub-percent and percent-level precision, respectively. This marks a substantial improvement over earlier reactor measurements, with \textsf{JUNO} reducing the uncertainty in $\Delta m^2_{21}$ to approximately $1.55\%$. For comparison, the \textsf{KamLAND}~\cite{KamLAND:2013rgu} experiment reported the solar mass-squared difference as
$\Delta m^2_{21} = (7.53^{+0.18}_{-0.18}) \times 10^{-5}\,\text{eV}^2$ corresponding to an uncertainty of approximately $2.81 \%$ while \textsf{KamLAND} data taken in 2013, from a combined three-flavored analysis of solar neutrino oscillation, reported the value of the solar mixing angle as   $\tan^2\theta_{12} = 0.436^{+0.029}_{-0.025}$ and $\Delta m^2_{21} = 7.53^{+0.18}_{-0.18} \times 10^{-5} \text{eV}^2$ \cite{KamLAND:2013rgu}. More recently, the phase-IV analysis of \textsf{Super-Kamiokande} (SK-IV)~\cite{Super-Kamiokande:2023jbt}, based on solar neutrino data, reported
\begin{equation}
\sin^2\theta_{12} = 0.306 \pm 0.013, \qquad
\Delta m^2_{21} = (6.10^{+0.95}_{-0.81}) \times 10^{-5}~\text{eV}^2,
\end{equation}
\begin{figure}
    \centering
    \includegraphics[width=0.8\linewidth]{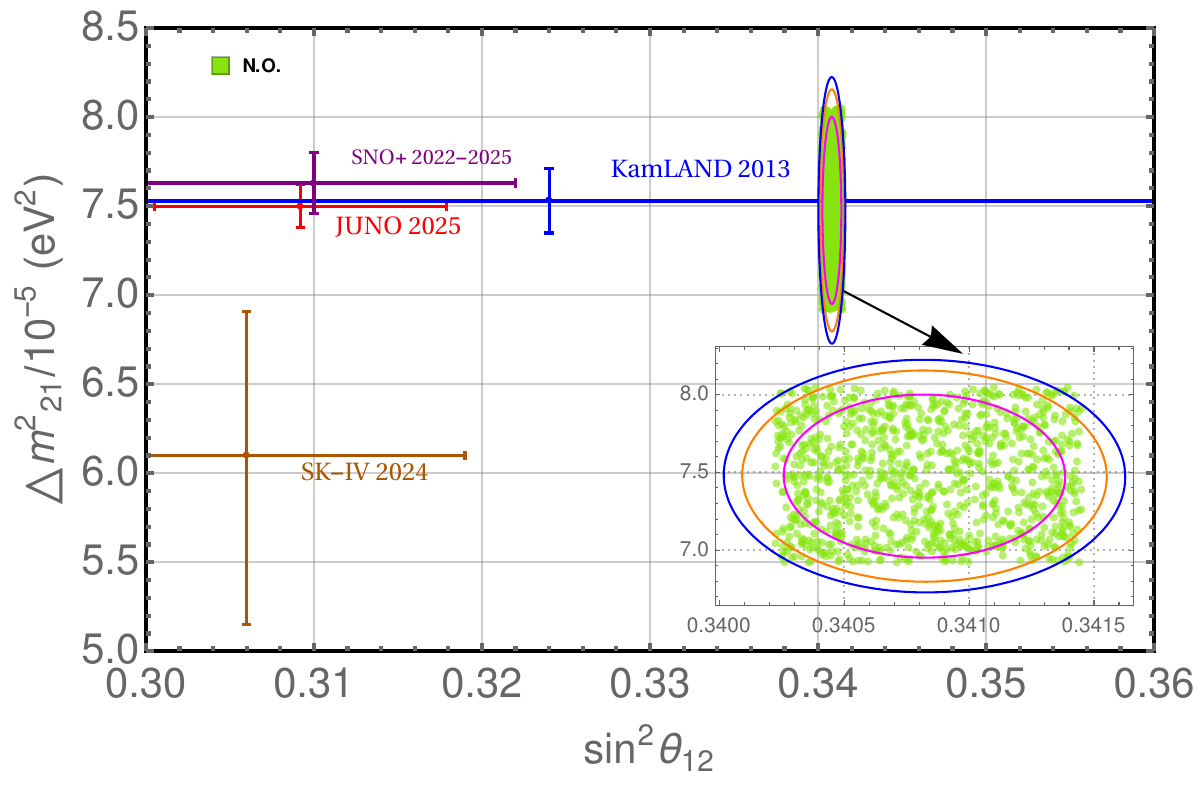}
    \caption{Correlation plot between measured value of $\sin^2\theta_{12}$ vs $\Delta m^2_{21}/10^{-5} \; \mbox{eV}^2$. The blue, orange and red contours in the magnified plot corresponds to $3\sigma, \; 2\sigma , \; 1\sigma$ allowed parameter space  of $\sin^2\theta_{12}$  vs $\Delta m^2_{21}/10^{-5} \; \text{eV}^2$ while the green points correspond to the double seesaw-invoked $A_4$ symmetry model prediction.}
    \label{fig:juno}
\end{figure}

Our model derived limits, prior to the modification of TM$_2$ was $\Delta m^2_{21} \in [6.92, 8.05] \times 10^{-5} \; \text{eV}^2$ align well with the experimental limits of JUNO (2025) KamLAND (2013), SK-IV limits (2024) and that for $\sin^2\theta_{12} \in [3.401, 3.413] $ falls within the limits ventured by the experiments KamLAND (2013), and NuFIT 6.0 neutrino oscillation analysis but is excluded by JUNO at more than $3\sigma$ . The correlation between the solar mixing parameter $\sin^2\theta_{12}$ and the mass-squared difference $\Delta m^2_{21}$ obtained from a scan over the viable parameter space of the present double seesaw-invoked $A_4$ framework is shown in Fig.\ref{fig:juno} displayed in green dotted points. The colored contours in the magnified inset correspond to the $1\sigma$ (red), $2\sigma$ (orange), and $3\sigma$ (blue) confidence regions, while the green points denote model solutions consistent with all theoretical and phenomenological constraints of the present framework. The superimposed experimental bands represent the recent precision determination from the JUNO 2025 projection, along with earlier measurements from KamLAND-2013~\cite{KamLAND:2013rgu} and Super-Kamiokande-IV (2024)~\cite{Super-Kamiokande:2023jbt}. The allowed region clearly favors normal mass ordering (N.O.), which emerges naturally in our framework as a consequence of the  $A_4$-symmetric structure of the light neutrino mass matrix and the hierarchical suppression inherent to the double seesaw mechanism. A pronounced correlation between $\sin^2\theta_{12}$ and $\Delta m^2_{21}$ is observed, reflecting the reduced parameter freedom imposed by the flavor symmetry. Importantly, the JUNO-preferred region of $\Delta m^2_{21}$ is well contained within the $1\sigma$–$2\sigma$ contours of our model, indicating excellent compatibility without fine-tuning. Given JUNO’s anticipated sub-percent sensitivity to both $\Delta m^2_{21}$ and $\sin^2\theta_{12}$, future data will decisively test this correlation, rendering the present double seesaw realization with $A_4$ symmetry highly predictive and potentially falsifiable in the solar neutrino sector.
\\
In addition, the SNO+ collaboration reported
$\Delta m^2_{21} = (7.63 \pm 0.17)\times 10^{-5}\;\text{eV}^2$ and
$\sin^2\theta_{12}=0.310\pm0.012$ from reactor antineutrino data collected between May~2022 and July~2025~\cite{SNO:2025chx}. It is also to be noted that Solar+KamLAND records $\Delta m^2_{21}/10^{-5} \approx 7.50 ^{+0.19}_{-0.18}, \; \sin^2\theta_{12}= 0.307 \pm 0.012 $ \cite{Super-Kamiokande:2023jbt} and Solar global records $\Delta m^2_{21}/10^{-5} \approx 6.10 ^{+0.95}_{-0.81}, \; \sin^2\theta_{12}= 0.303 \pm 0.013 $ \cite{Super-Kamiokande:2023jbt}. These data were recorded from October 2008 to May 2018 (2970 days) from the fourth phase of Super Kamiokande (SK IV)  for events of solar neutrinos energies of 3.49 - 19.49 meV. Most of these measurements are in excellent agreement with the model-predicted ranges, thereby reinforcing the robustness of the double seesaw–invoked $A_4$ framework.


\subsection{Beyond the Exact TM2 Limit: Compatibility with JUNO}
\label{sec:BeyondTM2}

\begin{figure}[htb!]
\centering
\includegraphics[width=0.8\linewidth]{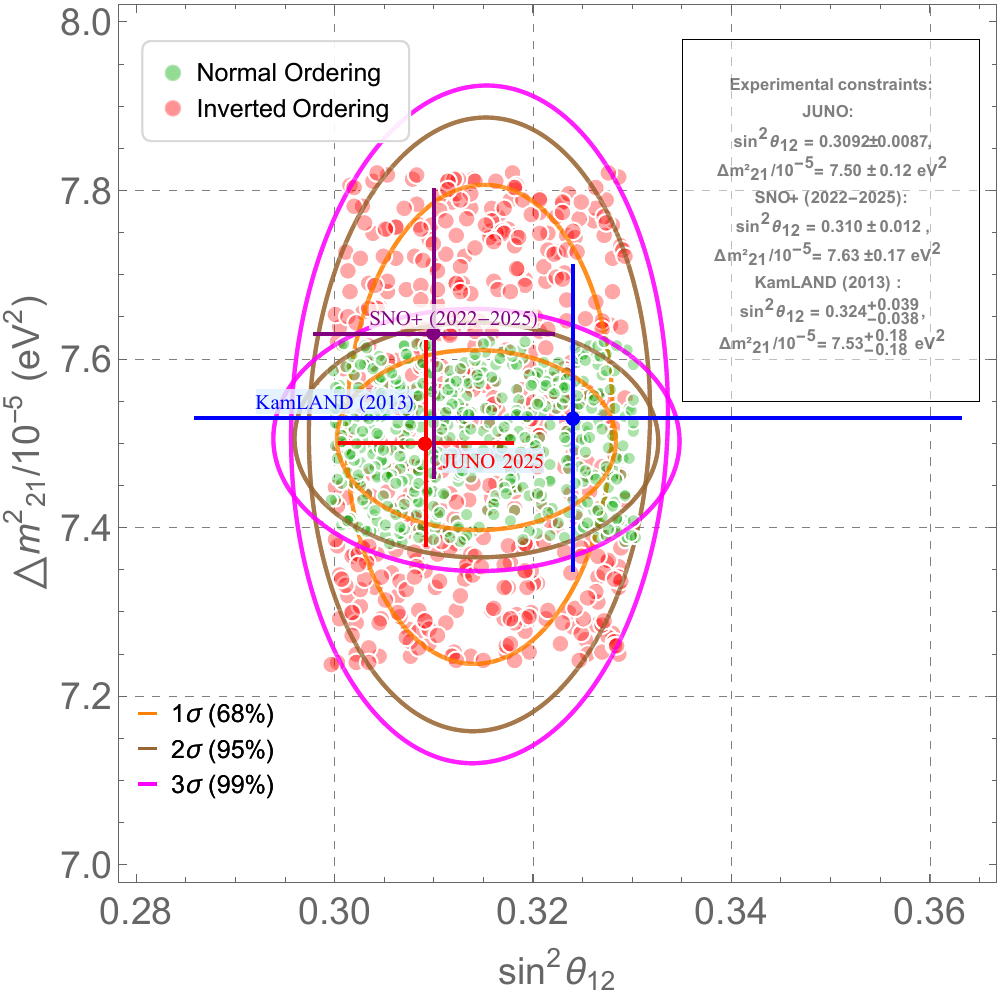}
\caption{Correlation between the predicted values of $\sin^2\theta_{12}$ and
$\Delta m_{21}^2/10^{-5}\,\mathrm{eV}^2$ in the generalized TM$_2$ mixing scenario
obtained by introducing perturbative corrections to the sterile Majorana mass matrix
$M_S$ within the double seesaw--invoked $A_4$ framework. The magenta, brown, and orange
contours in the magnified panel represent the experimentally allowed
$3\sigma$, $2\sigma$, and $1\sigma$ regions, respectively, in the
$\sin^2\theta_{12}$--$\Delta m_{21}^2/10^{-5}\,\mathrm{eV}^2$ plane, while the green
points denote the model predictions after the inclusion of the correction terms.}
    \label{fig:juno1}
\end{figure}
As discussed in Sec.~\ref{sec: beyond TM2}, the introduction of a small perturbation parameter $\varepsilon$ in the sterile-sector Majorana mass matrix generalizes the exact TM$_2$ mixing structure by relaxing the rigid correlation among the neutrino mixing angles. In particular, the modified TM$_2$ framework introduces an additional degree of freedom in the solar sector, allowing the solar mixing angle to deviate from the exact TM$_2$ prediction while preserving the successful description of the reactor and atmospheric mixing angles. Consequently, the well-known TM$_2$ sum rule, $\sin\theta_{12}=1/(\sqrt{3}\cos\theta_{13})$, is no longer exact, and the predicted value of the solar mixing angle can be shifted from $\sin^2\theta_{12} \simeq 0.341$ towards the experimentally preferred region. This extension is primarily motivated by the recent high-precision JUNO measurement, which favors $\sin^2\theta_{12} \simeq 0.304$ and is therefore in tension with the exact TM$_2$ prediction. 
\\
The impact of this generalized TM$2$ structure is illustrated in Fig.~\ref{fig:juno1}. The figure shows the correlation between the predicted values of $\sin^2\theta_{12}$ and $\Delta m_{21}^{2}$ after incorporating the perturbative corrections to the sterile-sector mass matrix $M_S$.  The green (red) points correspond to the model predictions for the normal (inverted) mass ordering, while the  magenta, brown, and orange contours represent the experimentally allowed $3\sigma$, $2\sigma$, and $1\sigma$ regions, respectively. In the exact TM$_2$ framework, the solar mixing angle is strongly correlated with the reactor mixing angle through the residual flavor symmetry, leading to the well-known prediction $\sin^2\theta_{12}\simeq0.341$. Although this prediction was compatible with earlier global analyses, it is now in clear tension with the recent high-precision JUNO measurement, which favors $\sin^2\theta_{12}\simeq0.304$. The introduction of the perturbation parameter $\varepsilon$ relaxes this rigid correlation by providing an additional degree of freedom in the solar sector, while preserving the successful description of the remaining oscillation observables. As illustrated in Fig.~\ref{fig:juno1}, the model-predicted points populate the experimentally allowed region in the $\sin^2\theta_{12}$--$\Delta m_{21}^2$ plane for both normal (green) and inverted (red) mass orderings, with the majority of the solutions lying within the $3\sigma$ contours and a significant fraction entering the $2\sigma$ and $1\sigma$ preferred regions. In particular, the predicted values span approximately
$0.30 \lesssim \sin^2\theta_{12}\lesssim0.33$ and
$\Delta m_{21}^2\simeq(7.4-7.6)\times10^{-5}\ {\rm eV}^2$, 
which are fully consistent with the latest JUNO determination and the current NuFIT~6.1 global analysis. Therefore, the generalized TM$_2$ structure generated by small perturbations in $M_S$ successfully restores compatibility with the recent solar-neutrino data while maintaining the predictive nature of the underlying $A_4$-symmetric double seesaw framework. This demonstrates that only a minimal departure from the exact residual-symmetry limit is sufficient to reconcile the model with the present oscillation data without altering its essential theoretical structure.
\begin{figure}[t]
\centering
\begin{subfigure}{0.48\textwidth}
\centering
\includegraphics[width=\linewidth]{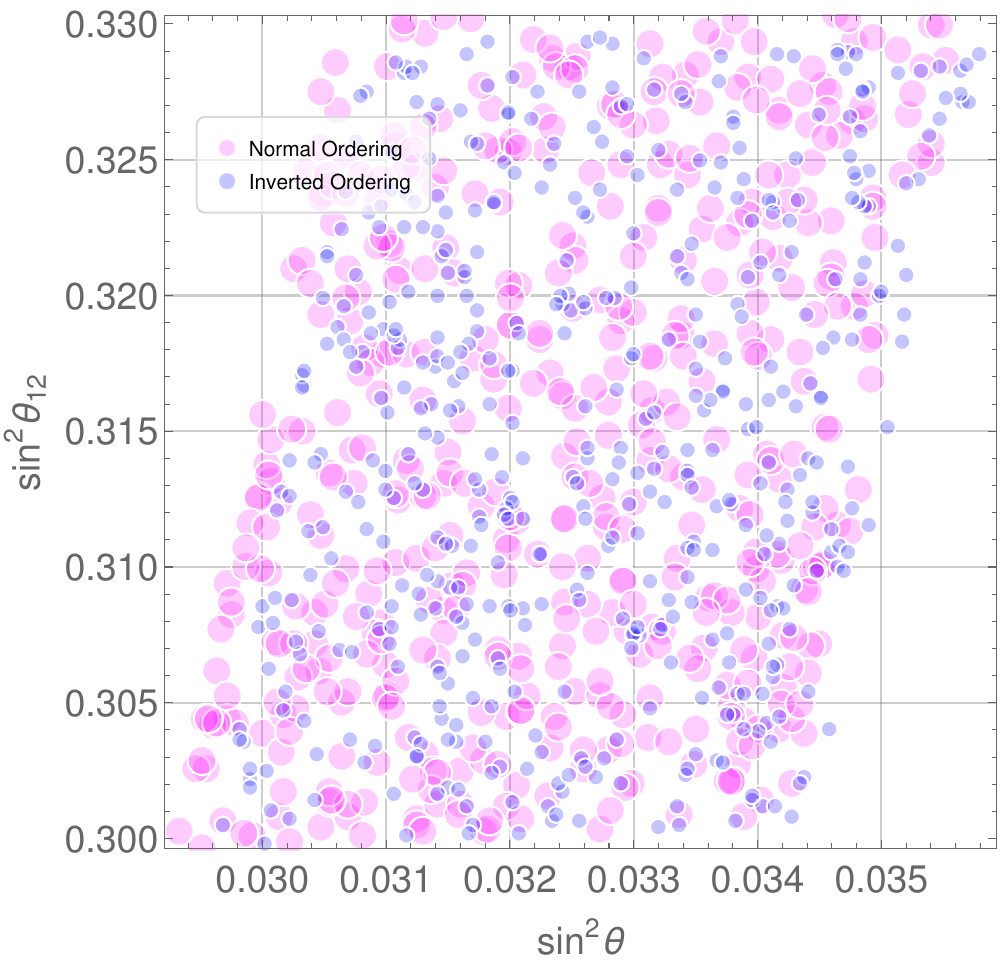}
\caption{$\sin^2\theta$ versus $\sin^2\theta_{12}$.}
\end{subfigure}
\hfill
\begin{subfigure}{0.48\textwidth}
\centering
\includegraphics[width=\linewidth]{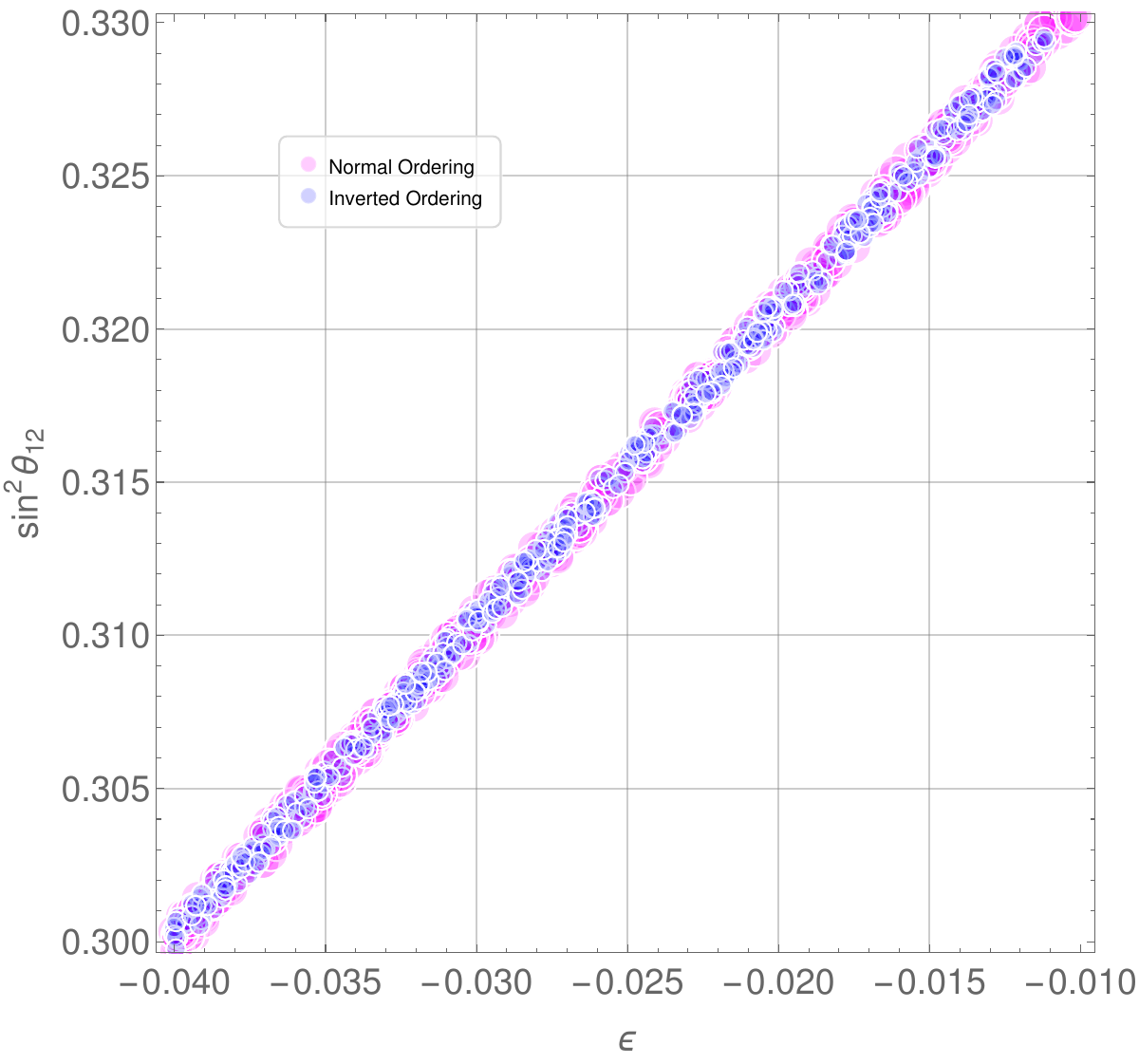}
\caption{$\varepsilon$ versus $\sin^2\theta_{12}$.}
\end{subfigure}
\caption{Correlations between the generalized TM$_2$ model parameters and the solar mixing angle. Panel (a) shows the dependence of $\sin^2\theta_{12}$ on the rotation angle $\theta$, while panel (b) illustrates its correlation with the perturbation parameter $\varepsilon$. Pink (blue) points correspond to the normal (inverted) mass ordering.}
\label{fig:corr1}
\end{figure}

\begin{figure}[t]
\centering
\begin{subfigure}{0.48\textwidth}
\centering
\includegraphics[width=\linewidth]{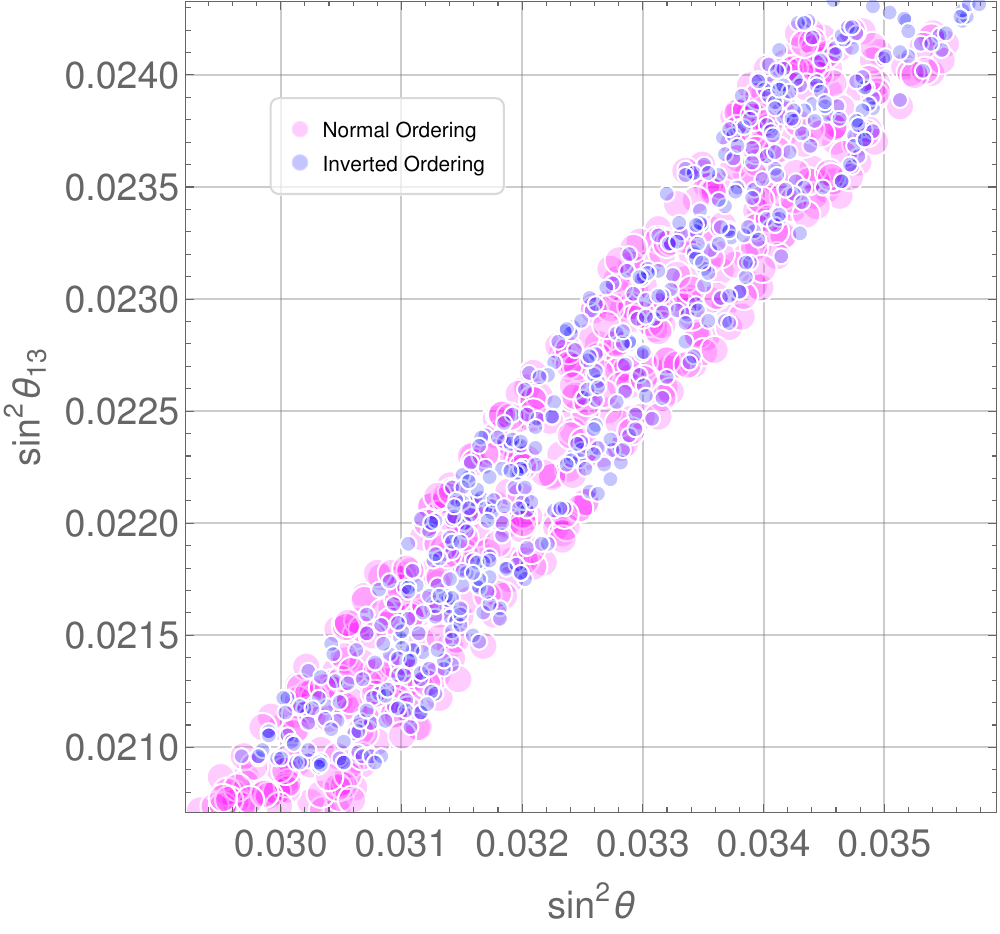}
\caption{$\sin^2\theta$ versus $\sin^2\theta_{13}$.}
\end{subfigure}
\hfill
\begin{subfigure}{0.48\textwidth}
\centering
\includegraphics[width=\linewidth]{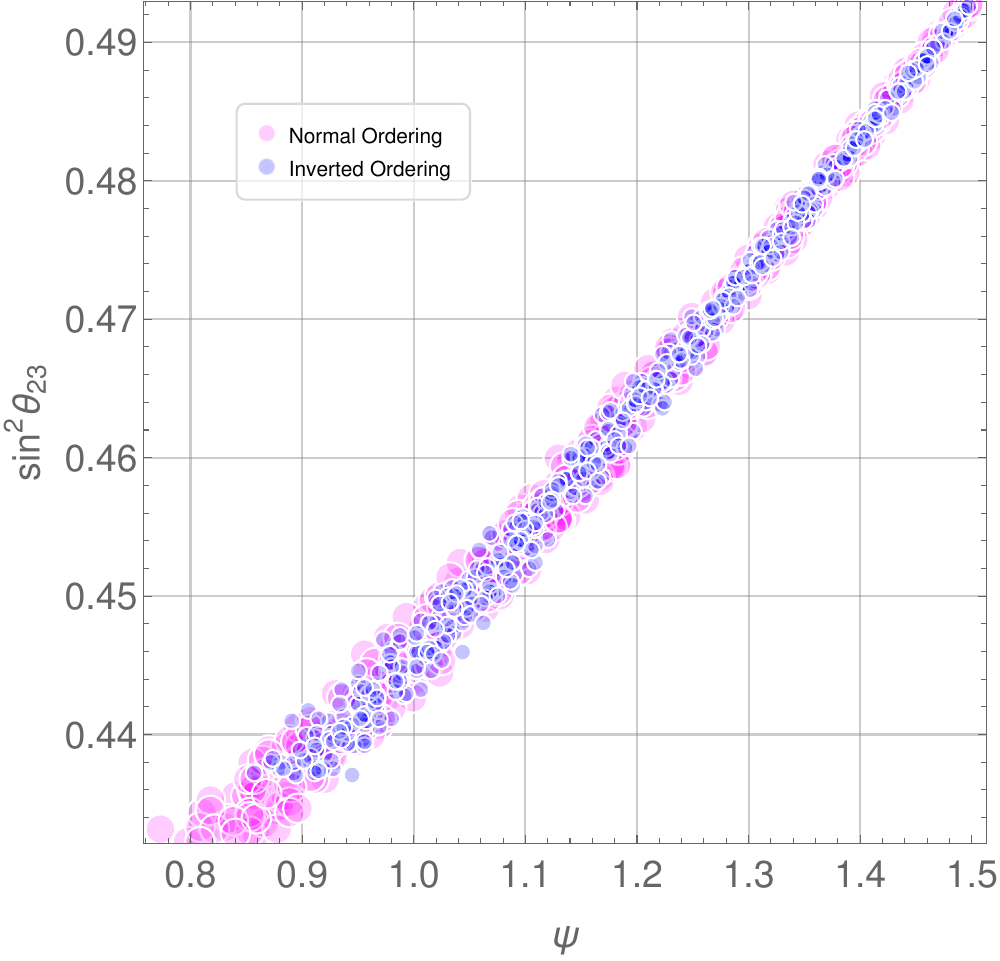}
\caption{$\psi$ versus $\sin^2\theta_{23}$.}
\end{subfigure}

\caption{Correlations of the generalized TM$_2$ parameters with the reactor and atmospheric mixing angles. Panel (a) shows the dependence of $\sin^2\theta_{13}$ on the rotation angle $\theta$, whereas panel (b) presents the correlation between the phase parameter $\psi$ and $\sin^2\theta_{23}$. Pink (blue) points denote the normal (inverted) ordering.}
\label{fig:corr2}
\end{figure}

To further understand the role of the generalized TM$_2$ parameters, we examine the correlations between the model parameters $(\theta,\psi,\varepsilon)$ and the neutrino oscillation observables. These correlations are shown in Figs.~\ref{fig:corr1}--\ref{fig:corr2} for both normal and inverted mass orderings. Figure~\ref{fig:corr1}(a) shows the correlation between the rotation parameter $\theta$ and the solar mixing angle $\sin^2\theta_{12}$. Unlike the exact TM$_2$ scenario, where $\sin^2\theta_{12}$ is tightly constrained by $\theta_{13}$, the introduction of the perturbation parameter $\varepsilon$ allows a considerably wider range of acceptable values,
$0.300 \lesssim \sin^2\theta_{12}\lesssim0.330$, 
while
$0.0295\lesssim\sin^2\theta\lesssim0.0358$. 
This demonstrates that the generalized TM$_2$ framework successfully accommodates the latest JUNO determination of the solar mixing angle. 
The dependence of the solar mixing angle on the perturbation parameter is illustrated in Fig.~\ref{fig:corr1}(b). A nearly linear correlation is observed between $\varepsilon$ and $\sin^2\theta_{12}$, consistent with Eq.~(\ref{eq:s12_modified}). The overlap of the normal- and inverted-ordering points indicates that this correlation is essentially insensitive to the neutrino mass ordering. 
Figure~\ref{fig:corr2}(a) presents the correlation between $\sin^2\theta$ and the reactor mixing angle. As expected from Eq.~(\ref{eq:s12_modified}), $\sin^2\theta_{13}$ is predominantly controlled by the rotation parameter $\theta$, resulting in a narrow linear correlation. This demonstrates that precise measurements of the reactor angle directly constrain the generalized TM$_2$ rotation parameter. The atmospheric sector is shown in Fig.~\ref{fig:corr2}(b), where $\sin^2\theta_{23}$ exhibits a strong correlation with the phase parameter $\psi$. The narrow diagonal band reflects the functional dependence predicted by Eq.~(\ref{eq:s23_modified}). Similar to the solar sector, the overlap between the normal- and inverted-ordering solutions indicates that this correlation is largely independent of the mass ordering.


\subsection{Impact on Atmospheric Neutrino Parameters from atmospheric data from Deep Core Ice-Cube and LBL experiments}
Atmospheric neutrino measurements further constrain the atmospheric neutrino  parameter space obtained from our model. The Deep Core Ice-Cube Neutrino Observatory  analyzed neutrinos with energies above 5 GeV.  Based on this 3387 days of data (2012-2021), Deep-Core Ice-Cube reports
$\Delta m^2_{32}=2.40^{+0.05}_{-0.04}\times10^{-3}\,\text{eV}^2$ and
$\sin^2\theta_{23}=0.54^{+0.04}_{-0.03}$~\cite{IceCubeCollaboration:2024ssx}. 
Similarly, the Super Kamiokande analyzed the data set taken from phase I to phase V (SK I-V) (from April 1996 to July 2020) and based on it reported that $\Delta m^2_{32} = 2.40^{+0.07}_{-0.09} \times 10^{-3}\; \text{eV}^2$, $\sin^2\theta_{23}=0.45^{+0.06}_{-0.05}$ and $\delta_{CP} = -1.75^{+0.76}_{-1.25} $ \cite{Super-Kamiokande:2023ahc}. Our model predicts the Dirac CP-violating phase to lie in the range
$\delta_{\rm CP}\in[0.77,\;1.50]~\mathrm{rad}$ for the normal mass ordering and
$\delta_{\rm CP}\in[0.86,\;1.50]~\mathrm{rad}$ for the inverted mass ordering.
The atmospheric mass-squared difference is obtained within
$\Delta m_{31}^{2}\in[2.45,\;2.59]\times10^{-3}\,\mathrm{eV}^{2}$ for the normal ordering, while the atmospheric mixing angle is predicted to lie in the ranges
$\sin^{2}\theta_{23}\in[0.4320,\;0.4932]$ for the normal ordering and
$\sin^{2}\theta_{23}\in[0.4370,\;0.4931]$ for the inverted ordering. These results indicate that, within the viable parameter space of the generalized TM$_2$ framework, the atmospheric mixing angle is preferred to lie in the lower octant for both mass orderings. The predicted ranges, summarized in Tables~\ref{tab:NO} and \ref{tab:IO}, are in good agreement with the current global neutrino oscillation data and remain fully compatible with the latest experimental measurements.

Long-baseline accelerator experiments provide complementary constraints on the atmospheric neutrino oscillation parameters. The recent NO$\nu$A analysis reports 
$\Delta m_{32}^{2}
=
2.431^{+0.036}_{-0.034}
\left(
-2.479^{+0.036}_{-0.036}
\right)
\times10^{-3}\ {\rm eV}^{2}$ 
for the normal (inverted) mass ordering, while the preferred value of the atmospheric mixing angle is
$\sin^{2}\theta_{23}=0.55^{+0.06}_{-0.02}$ for both orderings~\cite{NOvA:2025tmb}. The analysis exhibits a mild preference for the normal mass ordering, with a likelihood approximately 2.4 times larger than that for the inverted ordering. Furthermore, the combined T2K and NO$\nu$A analysis constrains the Dirac CP-violating phase to
$\delta_{\rm CP}\in[-1.38\pi,\;0.30\pi]$ for the normal ordering and
$\delta_{\rm CP}\in[-0.92\pi,\;-0.04\pi]$ for the inverted ordering, without establishing a significant preference for either mass ordering~\cite{T2K:2025wet}. The KM3NeT/ORCA atmospheric neutrino analysis also disfavors the inverted ordering and reports
$\sin^{2}\theta_{23}=0.51^{+0.04}_{-0.05}$ together with
$\Delta m_{31}^{2}=(2.18^{+0.25}_{-0.35})\times10^{-3}\ {\rm eV}^{2}$ for the normal ordering~\cite{KM3NeT:2024ecf}. While these experiments presently exhibit a mild preference for the higher octant of $\theta_{23}$, the octant has not yet been established conclusively. In contrast, the generalized TM$_2$ framework investigated here favors the lower-octant solution within the experimentally allowed parameter space, whereas the exact TM$_2$ realization naturally prefers the higher octant. Consequently, future precision measurements of the atmospheric mixing angle at DUNE and Hyper-Kamiokande will provide a powerful discriminator between the exact and generalized TM$_2$ scenarios.

\subsection{Neutrinoless Double Beta Decay Implications from JUNO data}
Neutrinoless double beta decay ($0\nu\beta\beta$) provides a unique probe of the Majorana nature of neutrinos and is directly sensitive to the effective Majorana mass parameter $m_{ee}$, which depends on the neutrino masses, Majorana phases, and the leptonic mixing angles, in particular $\theta_{12}$ and $\theta_{13}$. Consequently, any improvement in the precision of oscillation parameters has a direct and quantitatively significant impact on the allowed range of $m_{ee}$ for both normal (NO) and inverted (IO) mass orderings. 
The implications of the recent \textsf{JUNO} precision measurements extend beyond oscillation phenomenology and play a crucial role in constraining the effective Majorana mass parameter $m_{ee}$ relevant for neutrinoless double beta ($0\nu\beta\beta$) decay. This naturally leads us to the study of neutrinoless double beta decay ($0\nu\beta\beta$),
\begin{equation}
{}^{A}_{Z}N \;\rightarrow\; {}^{A}_{Z+2}N + 2e^-,
\end{equation}
a lepton-number-violating process in which no neutrinos are emitted and only two electrons appear in the final state. Such a decay can occur if neutrinos are Majorana particles, and its amplitude is governed by the effective Majorana mass parameter $m_{ee}$, defined as
\begin{subequations}
\begin{eqnarray}
\left| m_{ee} \right|&=&
  \bigg| |U^2_{e1}|\, m_1 + |U^2_{e2}|\, m_2 e^{i \alpha} + |U^2_{e3}|\, m_3 e^{i \beta} \bigg|  \\ &=&
  \left| c^2_s c^2_r m_1 + s^2_s c^2_r m_2 e^{i \alpha} + s^2_r m_3 e^{i \beta} \right| \,.
\label{eq:mee-std}
\end{eqnarray}
\end{subequations}
Here, $m_i$ are denoting the mass eigenvalues for light neutrinos while the Majorana phases are denoted as $\alpha, \beta$. Also, $U_{ei}$ represents the elements of PMNS mixing matrix and used the notation for mixing angles as $(c_\alpha, s_\alpha)\equiv (\cos\theta_\alpha, \sin\theta_\alpha)$ where 
$\theta_a\equiv \theta_{23}$ is the atmospheric mixing angle, $\theta_r\equiv \theta_{13}$ is the reactor mixing angle and $\theta_s\equiv \theta_{12}$ as the solar mixing angle. 

\begin{figure}[htb!]
    \centering
    \includegraphics[width=1\linewidth]{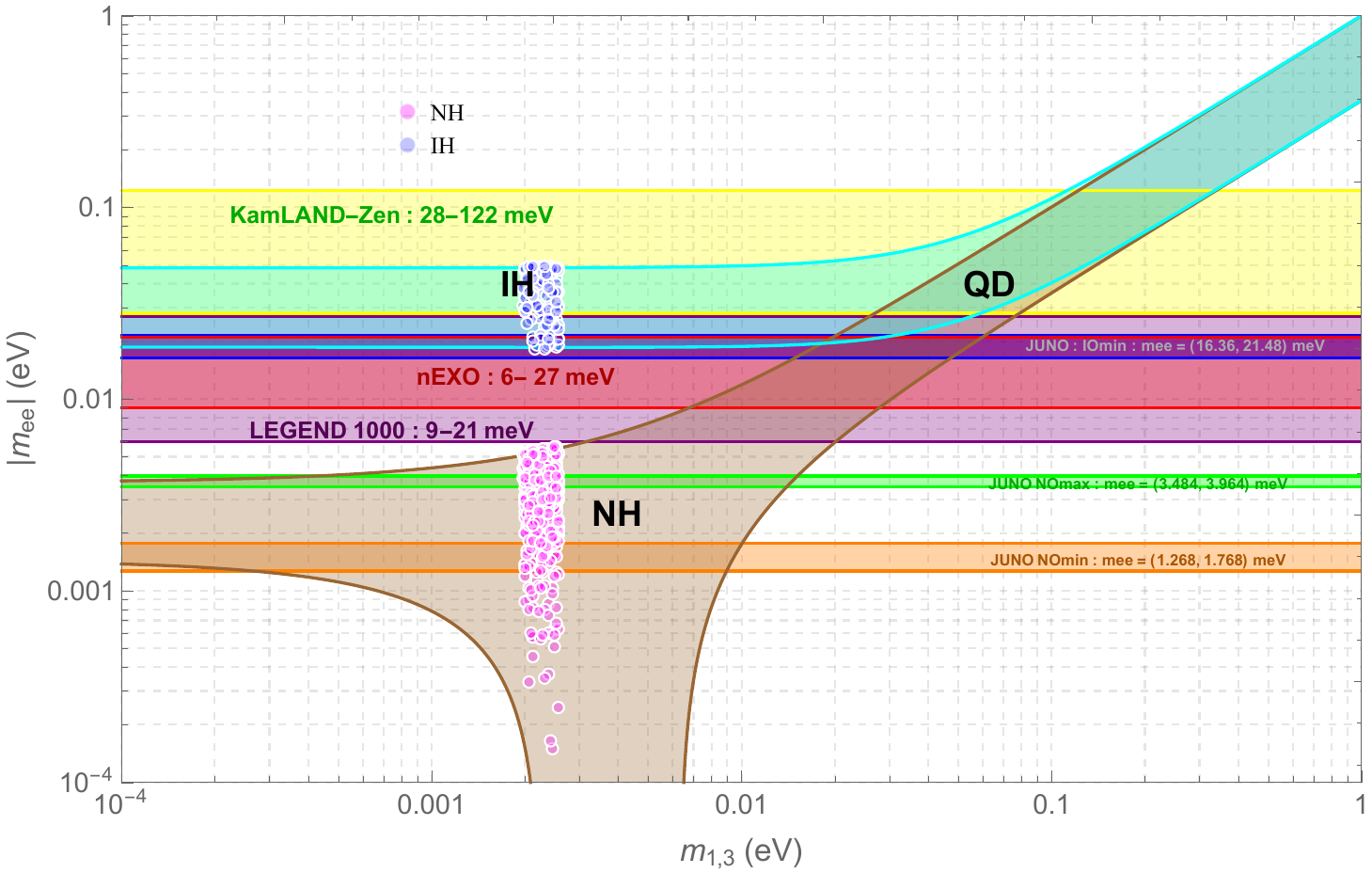}
    \caption{Effective Majorana mass parameter $m^{\nu}_{\rm ee} \equiv m_{\rm ee}$ with the variation of lightest  neutrino mass $m_1 \; (\mathrm{or} \; m_3)$ for normal ordering (NO) and inverted ordering (IO). The yellow horizontal band represents the experimental limits by \textsf{KamLAND-Zen}\,$(28-122\,\mathrm{meV})$~\cite{KamLANDZen_PRL2023,KamLAND-Zen:2024eml} and the red band corresponds to \textsf{nEXO}\, $(6-27\,\mathrm{meV})$~\cite{nEXO:2021ujk,nEXO:2017nam} and the purple band corresponds to \textsf{LEGEND 1000}\, $(9-21\,\mathrm{meV})$~\cite{LEGEND:2021bnm}  . The pink and blue points represents the allowed solutions obtained from our model for normal and inverted mass orderings respectively.}
    \label{fig:nubb}.
\end{figure}
Figure~\ref{fig:nubb} illustrates the allowed $3\sigma$ regions in the
$m_{ee}$--$m_{\rm lightest}$ plane obtained within the generalized TM$_2$
framework by varying the neutrino oscillation parameters over their allowed
ranges from the latest \textsf{NuFIT-6.1} global analysis~\cite{Esteban:2024eli},
together with the recent JUNO constraints. The light-brown and light-blue shaded
bands denote the experimentally allowed regions for the normal and inverted
mass orderings, respectively, while the pink (blue) points represent the
predictions of the double seesaw--invoked $A_4$ flavour model for the normal
(inverted) ordering. The figure demonstrates that the model predictions are
fully compatible with the current oscillation data and lie within the
experimentally allowed regions for both mass orderings. The overlapping region
between the normal and inverted ordering bands corresponds to the
quasi-degenerate neutrino mass spectrum, which is strongly constrained by the
latest cosmological observations and direct neutrino-mass measurements from 
\textsf{PLANCK} and \textsf{KATRIN}, respectively.  The translated limit on lightest neutrino mass has been derived from the experimental limit on sum of light neutrino masses and at present, the bound on the sum of neutrino masses from \textsf{PLANCK} is $\sum m_\nu < 0.12\,\text{eV}$ at 95\% C.L.~\cite{Planck:2018vyg}. 
The absolute neutrino mass scale can also be probed through direct kinematic measurements of the effective electron neutrino mass,
\begin{equation}
m_{\beta}=\sqrt{\sum_{i=1}^{3}|U_{ei}|^{2}m_{i}^{2}}. 
\end{equation}
The KATRIN experiment was originally designed to reach a sensitivity of
$m_{\beta}<0.2~\mathrm{eV}$ (90\% C.L.) after completion of its full data-taking program~\cite{KATRIN:2019yun}. Recently, based on 259 days of data collected during the 2019--2021 measurement campaigns, the KATRIN Collaboration reported the most stringent direct laboratory constraint on the effective electron neutrino mass,
\begin{equation}
m_{\beta}<0.45~\mathrm{eV}\qquad (90\%~\mathrm{C.L.}),
\end{equation}
providing the current best model-independent bound on the absolute neutrino mass scale~\cite{KATRIN:2024cdt}.
Because of the restricted structure of the light neutrino mass matrix enforced by the $A_4$ symmetry and the hierarchical suppression intrinsic to the double seesaw mechanism, the model predicts a narrow and correlated range of  $m_{ee} \in 0.290-5.58\; \text{meV} \; ( 16.20-49.12\; \text{meV})$ for NO (IO), fully consistent with the JUNO-updated $3\sigma$ intervals. The recent first data release from the JUNO experiment has led to a substantial reduction in the uncertainties of the solar mixing angle $\theta_{12}$ and the mass-squared difference $\Delta m^2_{21}$. This improved precision propagates into tighter constraints on $m_{ee}$, as explicitly demonstrated in a recent work~\cite{Ge:2025cky}. In particular, the relative uncertainties in the determination of $m_{ee}$ have been reduced to $22.0\%$ for $\langle m_{ee}\rangle^{\mathrm{IO}}_{\min}$, $22.5\%$ for $\langle m_{ee}\rangle^{\mathrm{NO}}_{\min}$, and $23.1\%$ for $\langle m_{ee}\rangle^{\mathrm{NO}}_{\max}$. As a result, the updated $3\sigma$ intervals inferred using JUNO first data are
\begin{align}
\langle m_{ee}\rangle^{\mathrm{IO}}_{\min} &= (16.36,\;21.48)\,\mathrm{meV}, \nonumber\\
\langle m_{ee}\rangle^{\mathrm{NO}}_{\min} &= (1.268,\;1.768)\,\mathrm{meV}, \nonumber\\
\langle m_{ee}\rangle^{\mathrm{NO}}_{\max} &= (3.484,\;3.964)\,\mathrm{meV}.
\end{align}
The present experimental limits on the effective Majorana mass are provided by \textsf{GERDA}\,$(79-118\,\mathrm{meV})$~\cite{GERDA:2020xhi}, \textsf{LEGEND 200}\,$(75-200\,\mathrm{meV}$\,\cite{LEGEND:2025jwu}) for $^{76}\mathrm{Ge}$ isotope, \textsf{EXO}~\cite{EXO-200:2019rkq}, \textsf{KamLAND-Zen}\,$(28-122\,\mathrm{meV}$~\cite{KamLAND-Zen:2024eml}) for $^{136}\mathrm{Xe}$ isotope, and \textsf{CUORE}\,$(70-240\,\mathrm{meV}$~\cite{CUORE:2019yfd}) for $^{130}\mathrm{Te}$ isotope.
Of particular importance is the fact that the upper range of $\langle m_{ee}\rangle_{\max}$ for normal ordering falls within the projected sensitivity of forthcoming next-generation $0\nu\beta\beta$ experiments. Experiments such as \textsf{nEXO} ($^{136}\mathrm{Xe}$)~\cite{nEXO:2021ujk,nEXO:2017nam} and \textsf{LEGEND-1000} ($^{76}\mathrm{Ge}$)~\cite{LEGEND:2021bnm} aim to reach sensitivities of $m_{ee} \simeq 6-27\,\mathrm{meV}$ and $m_{ee} \simeq 9-21\,\mathrm{meV}$, respectively, while \textsf{CUPID} targets a sensitivity of $m_{ee} \simeq 12-34\,\mathrm{meV}$~\cite{CUPID:2022wpt,CUPID:2020aow}. 

The model predicts narrow allowed regions for both mass orderings, consistent with current experimental limits from the \textsf{GERDA}, \textsf{KamLAND-Zen} and \textsf{LEGEND 200} experiments. Importantly, the predicted $m_{ee}$ ranges fall within the reach of next-generation experiments like \textsf{nEXO}, \textsf{LEGEND-1000}, offering a realistic opportunity to test the model. Finally, the presence of Majorana neutrino masses and CP-violating phases in the double seesaw–invoked $A_4$ framework naturally provides the ingredients required for leptogenesis, thereby linking neutrino properties with the origin of the baryon asymmetry of the Universe. The synergy between \textsf{JUNO}’s precision oscillation measurements and these upcoming $0\nu\beta\beta$ searches therefore renders the present double seesaw–invoked $A_4$ framework highly predictive and experimentally testable, with the potential to either confirm or decisively constrain the model in the near future. 
\subsection{Correlation between Dirac CP Phase and Jarlskog Invariant}
The Jarlskog invariant $J_{\rm CP}$ provides a rephasing-invariant measure of Dirac CP violation in the leptonic sector and is directly determined by the three leptonic mixing angles and the Dirac CP-violating phase $\delta$~\cite{Jarlskog:1985ht,Esteban:2020cvm,King:2013eh,Harrison:2002kp}. In this subsection, we evaluate the correlation between $J_{\rm CP}$ and the Dirac CP phase within the framework of the $A_4$-symmetric double seesaw model for both the normal (NO) and inverted (IO) neutrino mass orderings. 
The Jarlskog invariant is computed using the leptonic mixing angles $(\theta_{12},\,\theta_{13},\,\theta_{23})$ together with the Dirac CP-violating phase $\delta$. In the numerical analysis, the reactor mixing angle $\theta_{13}$ is taken as an input parameter within its allowed NuFIT~6.0 range. The remaining mixing parameters, namely $\theta_{12}$ and $\theta_{23}$, are then determined from Eqs.~(3.18) and (3.20), respectively, after obtaining the rotation parameter $\theta$ from Eq.~(3.19). This procedure allows us to evaluate the corresponding values of $J_{\rm CP}$ over the experimentally allowed parameter space.

\begin{figure}[t]
\centering
\begin{subfigure}{0.48\textwidth}
\centering
\includegraphics[width=\linewidth]{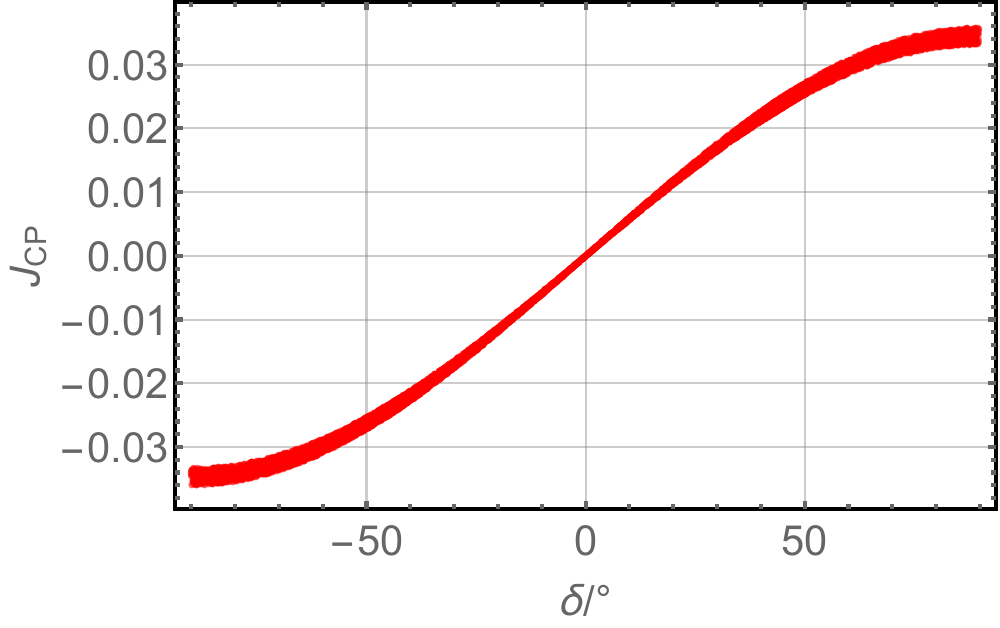}
\caption{}
\label{jcp.a}
\end{subfigure}
\hfill
\begin{subfigure}{0.48\textwidth}
\centering
\includegraphics[width=\linewidth]{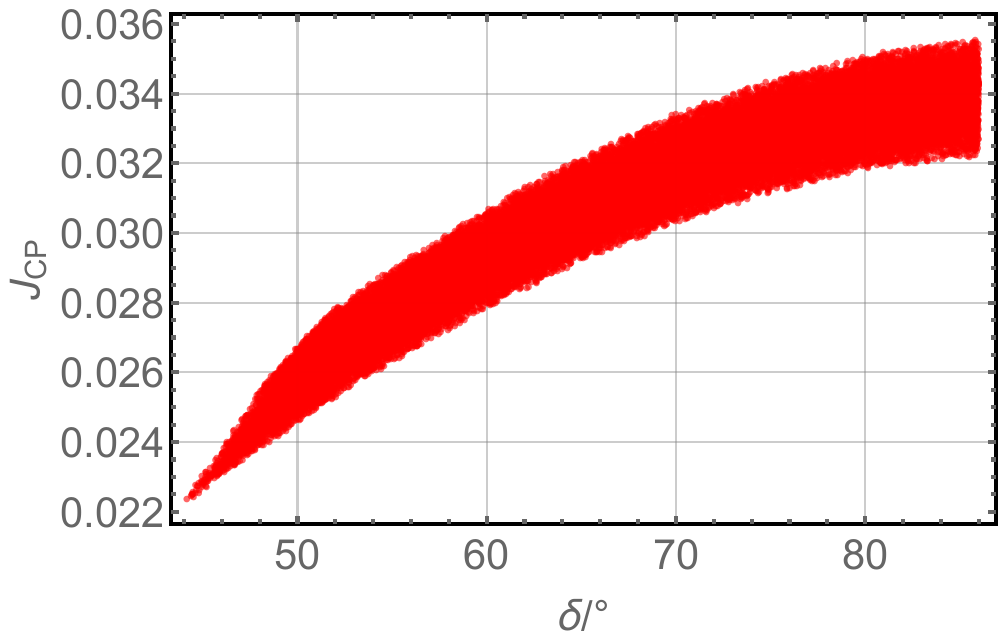}
\caption{}
\label{jcp.b}
\end{subfigure}

\caption{shows the correlation plots between the Dirac CP phase $\delta$ and the Jarlskog invariant $J_{\rm CP}$ for both normal mass ordering (NO) and inverted mass ordering (IO) cases, where in (a) the plot corresponds to $TM_2$  parametrization and in (b) the plot corresponds to the generalized $TM_2$ parametrization after inducing the perturbation parameter $\epsilon$ in it.}
\label{fig:Jcp_delta}
\end{figure}

As seen from Fig.~\ref{jcp.a}, the Jarlskog invariant exhibits a smooth and monotonic dependence on the Dirac CP phase and changes sign as $\delta$ crosses zero. 
The horizontal axis represents the Dirac CP phase $\delta$ (in degrees), while the vertical axis denotes the Jarlskog invariant $J_{\rm CP}$, which provides a rephasing-invariant measure of CP violation in the lepton sector.
This behavior follows directly from the general relation
\begin{equation}
J_{\rm CP} = \frac{1}{8}
\sin 2\theta_{12}\,
\sin 2\theta_{23}\,
\sin 2\theta_{13}\,
\cos \theta_{13}\,
\sin\delta,
\label{jcp}
\end{equation}
which holds for the standard parametrization of the PMNS matrix. Consequently, CP conservation is recovered for $\delta = 0^\circ$ and $\delta = \pm 180^\circ$, while maximal CP violation occurs near $\delta \simeq \pm 90^\circ$.
The predicted magnitude of the Jarlskog invariant in the present $A_4$ double seesaw framework lies in the range
\begin{equation}
|J_{\rm CP}| \simeq (3\text{--}4)\times 10^{-2},
\end{equation}
which is consistent with current global-fit results of neutrino oscillation data. Notably, the predictions corresponding to normal and inverted mass orderings largely overlap, indicating that the correlation between $\delta$ and $J_{\rm CP}$ is only weakly sensitive to the neutrino mass hierarchy. This feature originates from the dominant role played by the $A_4$-symmetric flavor structure of the mass matrices. 
Fig \ref{jcp.b} presents the Jarlskog invariant measured after the reparameterization of $U_{{\rm TM}_2}$, which also shows a monotonic dependence on the Dirac CP phase in the interval $J_{\rm CP} \; \in \; [0.024,0.035]$. This is in agreement with the latest NuFIT 6.1 data (2025), $J_{\rm CP} \in [-0.04,\; 0.04]$.
Thus, Fig.~\ref{fig:Jcp_delta} demonstrates that the $A_4$ flavor symmetric double seesaw model naturally accommodates sizable leptonic CP violation and predicts a well-defined correlation between the Dirac CP phase and the Jarlskog invariant. These predictions can be tested in current and upcoming long-baseline neutrino oscillation experiments such as T2K, NO$\nu$A, DUNE, and Hyper-Kamiokande.

The recent high-precision results from the JUNO experiment provide an important complementary test of the present framework by significantly improving the determination of the solar mixing angle $\theta_{12}$ and the mass-squared difference $\Delta m^2_{21}$, which directly enter the Jarlskog invariant~\cite{JUNO:2015zny,JUNO:2022mxj}. Using the JUNO-constrained parameter space, we find that the predicted range $|J_{\rm CP}|\simeq(3$--$4)\times10^{-2}$ remains unchanged for both normal and inverted mass orderings. This demonstrates that the $\delta$--$J_{\rm CP}$ correlation predicted by the $A_4$ flavor-symmetric double seesaw model is robust against the latest precision oscillation data. The synergy between JUNO and forthcoming long-baseline experiments will therefore provide a decisive test of leptonic CP violation in this framework.
\section{Conclusion}
\label{sec:conclusion}
In this work, we have presented a comprehensive study of the double seesaw mechanism embedded within an $A_4$ flavour-symmetric framework. By extending the Standard Model with three right-handed neutrinos and three gauge-singlet sterile fermions, we have shown that the interplay between the double seesaw mechanism and the $A_4$ flavour symmetry naturally leads to highly predictive structures for the Dirac neutrino mass matrix $M_D$, the right-handed--sterile mixing matrix $M_{RS}$, and the sterile-sector Majorana mass matrix $M_S$. Appropriate $A_4$ charge assignments together with simple flavon vacuum alignments generate a constrained neutrino mass matrix capable of reproducing the observed pattern of neutrino masses and flavour mixing.

At leading order, the model predicts the exact TM$_2$ mixing structure arising from a tribimaximal (TBM) pattern corrected by a single complex rotation in the $(1,3)$ sector. This framework successfully reproduces the observed reactor mixing angle while yielding characteristic correlations among the neutrino mixing parameters and CP-violating phases. However, the recent high-precision JUNO measurement of the solar mixing angle has significantly strengthened the experimental constraints on flavour models. In particular, the exact TM$_2$ prediction,  $\sin^2\theta_{12} \simeq 0.341$,  is found to be incompatible with the current JUNO-preferred value, thereby motivating a modification of the exact residual-symmetry structure.

To address this tension, we introduced a generalized TM$_2$ framework by incorporating a small perturbation into the sterile-sector Majorana mass matrix $M_S$. The additional perturbation parameter relaxes the rigid TM$_2$ sum rule and provides an extra degree of freedom in the solar sector, allowing the predicted value of $\sin^2\theta_{12}$ to vary within the experimentally preferred interval while preserving the successful predictions for the reactor and atmospheric mixing angles. We demonstrated that only a minimal departure from the exact residual-symmetry limit is sufficient to reconcile the model with the latest JUNO and NuFIT~6.1 oscillation data. A detailed numerical analysis was performed for both normal and inverted neutrino mass orderings by scanning the complete parameter space of the model. The generalized TM$_2$ framework successfully reproduces the observed neutrino mass spectrum and mixing parameters, predicts characteristic correlations among the CP-violating phases and oscillation observables, and yields viable parameter regions for neutrinoless double beta decay and leptogenesis. In particular, the correlation between the solar mixing angle and the solar mass-squared difference provides a distinctive prediction that can be tested with future high-precision oscillation measurements.

The generalized TM$_2$ framework also yields definite predictions for neutrinoless double beta decay. The effective Majorana mass parameter,
$m_{\beta\beta}$, obtained in the present $A_4$-symmetric double seesaw model is fully consistent with the current experimental limits for both normal and inverted mass orderings.  
Consequently, the combined precision of future oscillation measurements and next-generation $0\nu\beta\beta$ searches will provide a powerful probe of the present framework. 
In summary, the interplay between the double seesaw mechanism, the $A_4$ flavour symmetry, and the generalized TM$_2$ mixing structure provides a predictive and experimentally testable framework for understanding neutrino masses and flavour mixing. 

\section*{Acknowledgement}
SP acknowledges the financial support under MTR/2023/000687 funded by SERB, Govt. of India. 
Two of the authors (CD and SKN) gratefully acknowledge the Institute of Physics (IOP), 
Bhubaneswar for valuable discussions and support during the preparation of the manuscript. \\
MRD would like  to thank Maxim O. Gonchar (Dubna, JINR and Irkutsk State U.)  and  Liudmila Kolupaeva (Dubna, JINR) for giving permission to use the valuable information in their GitLab repository on the recent neutrino experimental findings. 
\appendix
\section{Lagrangian Construction and Mass Matrices in the Double Seesaw Framework}
\label{app:A4_DSM}
In this appendix, we present the detailed Yukawa Lagrangian and the resulting neutral fermion mass matrices of the double seesaw mechanism realized within the framework of the $A_4$ discrete flavour symmetry. These results provide the theoretical foundation for the analytical derivation of the neutrino masses, lepton mixing matrix, and CP-violating observables discussed in the main text. Our construction follows the standard approach of assigning appropriate $A_4$ representations to the lepton and scalar fields and generating the desired flavour structures through the vacuum alignment of flavon fields. This framework is motivated by the original $A_4$ flavour models and their subsequent developments in neutrino phenomenology~\cite{Altarelli:2005yx,Altarelli:2005yp,Ishimori:2010au,King:2013eh}. Numerous extensions of $A_4$ flavour symmetry in non-supersymmetric seesaw frameworks, including studies of neutrino masses, leptogenesis, dark matter, and flavour phenomenology, have subsequently been developed in the literature~\cite{Nanda:2025lem,Devi:2023vpe,RickyDevi:2024ijc,CentellesChulia:2025pyn,Kumar:2023moh,CentellesChulia:2023osj,Nomura:2019xsb,Nomura:2023usj,Nomura:2024nwh,Nomura:2024vus,Nomura:2022mgf,Kashav:2025mch,Nomura:2024atp,Singh:2024imk,Kashav:2022kpk,Kashav:2021zir,Priya:2025wdm,Mohanta:2023tzf,Mishra:2022tnf,Mishra:2022egy,Behera:2020lpd,Behera:2020sfe,Petcov:2024vph,Petcov:2023vws,Petcov:2022fjf,Novichkov:2018yse,Petcov:2018mvm,Petcov:2018snn,Zhang:2024rwv,Abbas:2022slb,Nagao:2021rio,deMedeirosVarzielas:2021pug,Wang:2019xbo,Kobayashi:2019gtp,Zhang:2019ngf,Kobayashi:2019xvz,Nomura:2019lnr,Nomura:2019jxj,Kobayashi:2018wkl,Baur:2024lcc,Chen:2024otk,Ding:2019gof,Ding:2019zxk}. In what follows, we summarize the field content, symmetry assignments, and the invariant Yukawa interactions that lead to the Dirac, right-handed neutrino, and sterile-sector Majorana mass matrices employed throughout this work.

\noindent
{\bf Charged-Lepton Sector:}\,
The Yukawa interaction responsible for charged-lepton masses is given by
\begin{align}
\mathcal{L}_{\ell}
&=
y^{ee}_{\ell}\,\overline{L}_{e_L} H e_R
+
y^{\mu\mu}_{\ell}\,\overline{L}_{\mu_L} H \mu_R
+
y^{\tau\tau}_{\ell}\,\overline{L}_{\tau_L} H \tau_R
+\text{h.c.},
\label{eq:Lag_charged}
\end{align}
where the left-handed lepton doublets transform as
${1}$, ${1}'$ and ${1}''$ under $A_4$, while the right-handed charged
leptons are singlets. After electroweak symmetry breaking, the charged-lepton
mass matrix is diagonal,
\begin{equation}
M_\ell
=
v
\begin{pmatrix}
y^{ee}_{\ell} & 0 & 0 \\
0 & y^{\mu\mu}_{\ell} & 0 \\
0 & 0 & y^{\tau\tau}_{\ell}
\end{pmatrix}
=
\begin{pmatrix}
m_e & 0 & 0 \\
0 & m_\mu & 0 \\
0 & 0 & m_\tau
\end{pmatrix},
\label{eq:Mell}
\end{equation}
implying a trivial charged-lepton diagonalization matrix,
$U_\ell=\mathbf{I}$.

\noindent
{\bf Double Seesaw Lagrangian with $A_4$ Symmetry:}\,
The relevant leptonic mass terms of the double seesaw mechanism are described by
the Lagrangian
\begin{align}
\mathcal{L}_{\text{DSM}}
&=
-\overline{L}_L \tilde{H} Y_D N_R
-\overline{N_R} M_{RS} S_L
-\frac{1}{2}\,\overline{S_L^c} M_S S_L
+\text{h.c.},
\label{eq:Lag_DSM}
\end{align}
where $N_R$ denotes the right-handed neutrinos transforming as an $A_4$ triplet,
$S_L$ represents sterile fermions, and $\tilde{H}=i\sigma_2 H^*$.

\noindent
{\bf Dirac Neutrino Mass Matrix $M_D$:}\, 
Due to the $A_4$ charge assignments, the Dirac Yukawa coupling
$\overline{L_L}\tilde{H}N_R$ is diagonal, yielding
\begin{equation}
M_D
=
v
\begin{pmatrix}
\alpha_D & 0 & 0 \\
0 & \beta_D & 0 \\
0 & 0 & \gamma_D
\end{pmatrix}.
\label{eq:MD}
\end{equation}
In the degenerate Yukawa limit considered in the main text, one has
$\alpha_D=\beta_D=\gamma_D\equiv \kappa_D$.

\noindent
{\bf Right-Handed--Sterile Mixing Matrix $M_{RS}$:}\, 
The mixing between $N_R$ and $S_L$ is described by
\begin{equation}
M_{RS}
=
v
\begin{pmatrix}
\alpha_{RS} & 0 & 0 \\
0 & \beta_{RS} & 0 \\
0 & 0 & \gamma_{RS}
\end{pmatrix},
\label{eq:MRS}
\end{equation}
where $v_T$ is the vacuum expectation value associated with the sterile sector.
In the degenerate limit, $\alpha_{RS}=\beta_{RS}=\gamma_{RS}\equiv \kappa_{RS}$.

\noindent
{\bf Sterile-Sector Majorana Mass Matrix $M_S$:}\,
The sterile-sector Majorana mass matrix is chosen to consist of an $A_4$-symmetric
leading term and a controlled symmetry-breaking correction,
\begin{align}
M_S &= M_S^{(0)} + M_S^{(\text{corr})},
\end{align}
with
\begin{equation}
M_S^{(0)}
=
\begin{pmatrix}
b+\tfrac{2a}{3} & -\tfrac{a}{3} & -\tfrac{a}{3} \\
-\tfrac{a}{3} & \tfrac{2a}{3} & b-\tfrac{a}{3} \\
-\tfrac{a}{3} & b-\tfrac{a}{3} & \tfrac{2a}{3}
\end{pmatrix},
\qquad
M_S^{(\text{corr})}
=
\begin{pmatrix}
0 & 0 & d \\
0 & d & 0 \\
d & 0 & 0
\end{pmatrix}.
\label{eq:MS}
\end{equation}
The leading term $M_S^{(0)}$ is exactly diagonalized by the tribimaximal mixing
matrix $U_{\text{TBM}}$, while the correction $M_S^{(\text{corr})}$ induces a
nontrivial $1$--$3$ rotation. This structure is responsible for generating a
nonzero reactor mixing angle and leptonic CP violation.

\noindent
{\bf Effective Light Neutrino Mass Matrix:}\,
After integrating out the heavy degrees of freedom, the effective light neutrino
mass matrix is given by the double seesaw formula
\begin{equation}
m_\nu
=
M_D\, (M_{RS}^{-1})^{T}\, M_S\, M_{RS}^{-1} M_D^{T}.
\label{eq:double_seesaw}
\end{equation}
In the degenerate Yukawa limit, this expression reduces to
\begin{equation}
m_\nu
=
\kappa_\nu\, M_S,
\qquad
\kappa_\nu
=
\left(\frac{v^2}{v_T^2}\right)
\left(\frac{\kappa_D^2}{\kappa_{RS}^2}\right),
\end{equation}
which makes explicit that all flavor structure originates from $M_S$, while the
overall neutrino mass scale is governed by the single parameter $\kappa_\nu$.
This decomposition provides the foundation for the analytic diagonalization and
phenomenological analysis presented in the main text.
\section{\texorpdfstring{Detailed Derivation of Mass and Mixing Structure in the Double Seesaw Framework with $A_4$ Symmetry}{Detailed Derivation of Mass and Mixing Structure in the Double Seesaw Framework with A4 Symmetry}}
\label{sec:MS-diagonalization}

In this section, we present a detailed analytic derivation of the mass eigenvalues,
mixing angles, and phases arising from the sterile-sector Majorana mass matrix
$M_S$ in the double seesaw framework endowed with an $A_4$ flavor symmetry.
The construction is such that the flavor structure of $M_S$ naturally separates
into an $A_4$-symmetric leading term and a controlled symmetry-breaking
correction, allowing for a transparent diagonalization procedure.

\subsection*{Structure of the sterile-sector Majorana mass matrix}

In the $A_4$-symmetric realization of the model, the sterile-sector Majorana mass
matrix can be written as
\begin{align}
M_S &= M_S^{(0)} + M_S^{(\mathrm{corr})},
\label{eq:MS_decomp-c}
\end{align}
with
\begin{equation}
M_S^{(0)} =
\begin{pmatrix}
b+\tfrac{2a}{3} & -\tfrac{a}{3} & -\tfrac{a}{3} \\
-\tfrac{a}{3} & \tfrac{2a}{3} & b-\tfrac{a}{3} \\
-\tfrac{a}{3} & b-\tfrac{a}{3} & \tfrac{2a}{3}
\end{pmatrix},
\qquad
M_S^{(\mathrm{corr})} =
\begin{pmatrix}
0 & 0 & d \\
0 & d & 0 \\
d & 0 & 0
\end{pmatrix}.
\label{eq:MS_terms}
\end{equation}
The leading term $M_S^{(0)}$ arises from $A_4$ triplet contractions of the type
$\mathbf{3}\otimes\mathbf{3}\to\mathbf{1}$ and is exactly diagonalized by the
tribimaximal mixing matrix $U_{\mathrm{TBM}}$. The correction term
$M_S^{(\mathrm{corr})}$ softly breaks the $A_4$ symmetry and introduces
off-diagonal entries that generate deviations from exact tribimaximal mixing.

Since in the degenerate Yukawa limit one has $M_D\propto M_{RS}\propto\mathbf{I}$,
the same unitary transformation that diagonalizes $M_S$ also diagonalizes the
effective light neutrino mass matrix (up to an overall scaling factor).
Therefore, the diagonalization of $M_S$ fully determines the neutrino mixing
structure.

\subsection*{Transformation to the TBM basis}

Transforming $M_S$ to the TBM basis, we obtain
\begin{align}
M_S' &\equiv U_{\mathrm{TBM}}^{T} M_S U_{\mathrm{TBM}} \nonumber \\
&=
\begin{pmatrix}
a+b-\tfrac{d}{2} & 0 & -\tfrac{\sqrt{3}}{2}\,d \\
0 & b+d & 0 \\
-\tfrac{\sqrt{3}}{2}\,d & 0 & a-b+\tfrac{d}{2}
\end{pmatrix}.
\label{eq:MS_TBM_basis}
\end{align}
The transformed matrix exhibits a nontrivial structure only in the $1$--$3$
block, while the $(2,2)$ entry remains diagonal. Consequently, the full
diagonalization requires only a single rotation in the $1$--$3$ plane.

\subsection*{Diagonalization via a single $1$--$3$ rotation}

The matrix $M_S'$ can be diagonalized by a unitary rotation
$U_{13}(\theta,\psi)$, such that
\begin{align}
M_S^{\mathrm{diag}}
&=
U_{13}^{T} M_S' U_{13}
=
(U_{13} U_{\mathrm{TBM}})^{T} M_S (U_{\mathrm{TBM}} U_{13}) \nonumber \\
&=
\mathrm{diag}\!\left(
M_{S_1} e^{i\phi_1},\;
M_{S_2} e^{i\phi_2},\;
M_{S_3} e^{i\phi_3}
\right).
\label{eq:MS_diag}
\end{align}
Requiring the off-diagonal $(1,3)$ element to vanish yields the condition
\begin{equation}
\tan 2\theta
=
\frac{2\sqrt{2}\,(a-b)}{a+b-2d},
\label{eq:tan2theta_MS}
\end{equation}
which determines the mixing angle associated with the $1$--$3$ rotation.

\subsection*{Eigenvalues of the sterile-sector mass matrix}

The resulting eigenvalues of $M_S$ are given by
\begin{align}
M_{S_1} &= a+\sqrt{b^2+d^2-bd}, \nonumber \\
M_{S_2} &= b+d, \nonumber \\
M_{S_3} &= a-\sqrt{b^2+d^2-bd},
\label{eq:MS_eigenvalues}
\end{align}
where the associated phases $\phi_{1,2,3}$ arise from the complex nature of the
parameters $a$, $b$, and $d$.

To make the phase dependence explicit, we introduce the dimensionless ratios
\begin{equation}
\frac{d}{b} = \lambda_1 e^{i\phi_{db}},
\qquad
\frac{a}{b} = \lambda_2 e^{i\phi_{ab}},
\label{eq:lambda_def}
\end{equation}
with $\phi_{ab}=\phi_a-\phi_b$ and $\phi_{db}=\phi_d-\phi_b$. In terms of these
parameters, the eigenvalues can be written as
\begin{align}
M_{S_1} &= b\!\left[
\lambda_2 e^{i\phi_{ab}}
+
\sqrt{1+\lambda_1^2 e^{2i\phi_{db}}
      -\lambda_1 e^{i\phi_{db}}}
\right], \nonumber \\
M_{S_2} &= b\!\left(1+\lambda_1 e^{i\phi_{db}}\right), \nonumber \\
M_{S_3} &= b\!\left[
\lambda_2 e^{i\phi_{ab}}
-
\sqrt{1+\lambda_1^2 e^{2i\phi_{db}}
      -\lambda_1 e^{i\phi_{db}}}
\right].
\label{eq:MS_eigenvalues_lambda}
\end{align}

This analytic structure demonstrates explicitly how the sterile-sector masses,
mixing angles, and CP-violating phases are governed by the $A_4$ symmetry and its
controlled breaking. The presence of a single complex rotation explains the
origin of nonzero reactor mixing and leptonic CP violation in a minimal and
predictive manner, providing the foundation for the phenomenological analysis
presented in the main text.

\section{Flavon Potential and Vacuum Alignment}
\label{sec:vacuum-alignment}

In this section, we discuss the scalar potential responsible for the spontaneous
breaking of the $A_4$ flavor symmetry and the resulting vacuum alignments that
lead to predictive mass textures in the double seesaw framework. The flavon
fields $\phi_T$, $\phi_S$, $\xi$, and $\xi'$ play distinct and complementary
roles in shaping the charged-lepton and neutrino mass structures. In particular,
the alignment of $\phi_T$ enforces a diagonal structure for the Dirac mass matrix
$M_D$ and the right-handed--sterile mixing matrix $M_{RS}$, while the combined
alignments of $\phi_S$, $\xi$, and $\xi'$ determine the flavor structure of the
sterile-sector Majorana mass matrix $M_S$. Since, in the degenerate Yukawa limit,
the effective light neutrino mass matrix satisfies $m_\nu=\kappa_\nu M_S$, the
entire flavor structure of light neutrinos is governed by the vacuum alignment
in the sterile sector~\cite{Altarelli:2005yp,Altarelli:2005yx,Ma:2001dn,Grimus:2013tva,deMedeirosVarzielas:2025byb}.

The most general renormalizable scalar potential invariant under
$A_4\times SU(2)_L\times U(1)_Y$ can be written as
\begin{equation}
V = V_T + V_S + V_\xi + V_{\xi'} + V_{\mathrm{mix}},
\label{eq:V_total}
\end{equation}
where each term corresponds to self-interactions or mutual interactions among
the flavon fields, as discussed below.

\subsection*{Triplet Potential for $\phi_T$}

The $A_4$-invariant scalar potential for the flavon triplet $\phi_T$ is given by
\begin{equation}
V_T =
\mu_T^2(\phi_T\!\cdot\!\phi_T)
+ \lambda_1(\phi_T\!\cdot\!\phi_T)^2
+ \lambda_2(\phi_T\!\cdot\!\phi_T)_{1'}
          (\phi_T\!\cdot\!\phi_T)_{1''}
+ \lambda_3(\phi_T\!\cdot\!\phi_T)_{3_s}
          \!\cdot\!(\phi_T\!\cdot\!\phi_T)_{3_s},
\label{eq:VT}
\end{equation}
where $(\cdots)$ denotes contractions into the corresponding $A_4$
representations. Minimization of $V_T$ yields
\begin{equation}
\frac{\partial V_T}{\partial \phi_{T1}} = 0
\quad \Rightarrow \quad
v_T^2 = -\frac{\mu_T^2}{2\lambda_1},
\end{equation}
with $\partial V_T/\partial\phi_{T2,3}=0$. The resulting vacuum alignment is
\begin{equation}
\langle \phi_T \rangle = (v_T,\,0,\,0),
\label{eq:vev_phiT}
\end{equation}
which preserves the $Z_3$ subgroup of $A_4$. This alignment ensures that both the
Dirac neutrino mass matrix $M_D$ and the right-handed--sterile mixing matrix
$M_{RS}$ are diagonal, thereby eliminating any source of flavor mixing from
these sectors.

\subsection*{Triplet Potential for $\phi_S$}

The triplet flavon $\phi_S$ governs the neutrino mass sector. Its most general
$A_4$-invariant potential is
\begin{equation}
V_S =
\mu_S^2(\phi_S\!\cdot\!\phi_S)
+ \kappa_1(\phi_S\!\cdot\!\phi_S)^2
+ \kappa_2(\phi_S\!\cdot\!\phi_S)_{1'}
           (\phi_S\!\cdot\!\phi_S)_{1''}
+ \kappa_3(\phi_S\!\cdot\!\phi_S)_{3_s}
           \!\cdot\!(\phi_S\!\cdot\!\phi_S)_{3_s}.
\label{eq:VS}
\end{equation}
Minimization yields
\begin{equation}
v_S^2 = -\frac{\mu_S^2}{2(\kappa_1+\kappa_3)},
\end{equation}
leading to the democratic alignment
\begin{equation}
\langle \phi_S \rangle = (v_S,\,v_S,\,v_S).
\label{eq:vev_phiS}
\end{equation}
This alignment preserves a residual $Z_2$ symmetry and generates the
tribimaximal structure of the sterile-sector mass matrix $M_S^{(0)}$, which in
turn determines the leading-order structure of the light neutrino mass matrix.

\subsection*{Singlet Flavons $\xi$ and $\xi'$}

The singlet flavons $\xi$ and $\xi'$ provide additional contributions to the
neutrino sector through their couplings to $\phi_S$. Their self-interaction
potentials are
\begin{equation}
V_\xi = \mu_\xi^2\xi^2+\lambda_\xi\xi^4,
\qquad
V_{\xi'} = \mu_{\xi'}^2\xi'^2+\lambda_{\xi'}\xi'^4.
\end{equation}
Minimization yields the vacuum expectation values
\begin{equation}
\langle \xi \rangle = u,
\qquad
\langle \xi' \rangle = u'.
\end{equation}
While $\xi$ contributes to the leading TBM structure of $M_S$, the singlet
$\xi'$ induces controlled symmetry-breaking corrections, generating deviations
from exact tribimaximal mixing and enabling a nonzero reactor angle $\theta_{13}$.

\subsection*{Mixed Interaction Terms}

The mixed flavon interactions are described by
\begin{equation}
V_{\mathrm{mix}} =
\alpha(\phi_T\!\cdot\!\phi_T)(\phi_S\!\cdot\!\phi_S)
+ \beta(\phi_T\!\cdot\!\phi_T)\xi^2
+ \gamma_3(\phi_S\!\cdot\!\phi_S)\xi^2
+ \gamma_4(\phi_S\!\cdot\!\phi_S)\xi'^2
+ \gamma_5\xi^2\xi'^2,
\end{equation}
which stabilize the vacuum configuration and correlate the magnitudes of the
VEVs without altering the alignment directions.

\subsection*{Summary of Vacuum Alignments}

The minimization of the full scalar potential leads to the vacuum structure
\begin{equation}
\langle \phi_T \rangle = (v_T,0,0), \quad
\langle \phi_S \rangle = (v_S,v_S,v_S), \quad
\langle \xi \rangle = u, \quad
\langle \xi' \rangle = u'.
\end{equation}
These alignments and their roles in the double seesaw framework are summarized in
Table~\ref{tab:vevs-summary}.

\begin{table}[htb!]
\centering
\caption{Flavon vacuum alignments and their roles in the double seesaw framework.}
\label{tab:vevs-summary}
\begin{tabular}{lcc}
\hline
\textbf{Flavon} & \textbf{VEV Alignment} & \textbf{Role} \\
\hline
$\phi_T$ & $(v_T,0,0)$ & Enforces diagonal $M_D$ and $M_{RS}$ \\
$\phi_S,\xi$ & $(v_S,v_S,v_S),\,u$ & Generates TBM structure of $M_S$ and $m_\nu$ \\
$\xi'$ & $u'$ & Provides controlled corrections to $M_S$ and $m_\nu$ \\
\hline
\end{tabular}
\end{table}

In summary, the vacuum alignment enforced by the $A_4$ symmetry ensures that all
lepton flavor mixing originates exclusively from the sterile-sector Majorana
mass matrix $M_S$. This feature renders the double seesaw framework highly
predictive: the charged-lepton, Dirac, and right-handed--sterile sectors remain
flavor diagonal, while the observed neutrino mixing and CP violation emerge
solely from the structure and controlled breaking of $M_S$.

\section{Detailed Derivation of Mixing Angles and CP Phases}
\label{app:mixing-derivation}

In this appendix, we present the detailed analytic derivation of the neutrino
mixing angles and CP-violating phases arising in the double seesaw--invoked
$A_4$ framework. The purpose of this section is to provide a transparent
derivation of the relations quoted in the main text.

\subsection*{Diagonalization of the $1$--$3$ Subspace}

After transforming the sterile-sector mass matrix into the tribimaximal basis,
the effective matrix in the $1$--$3$ subspace takes the form
\begin{equation}
M =
\begin{pmatrix}
X & Y\\
Y & Z
\end{pmatrix},
\label{eq:M13}
\end{equation}
where
\begin{equation}
X = a+b-\frac{d}{2}, \qquad
Y = \frac{\sqrt{3}}{2}\,d, \qquad
Z = a-b+\frac{d}{2}.
\end{equation}

This matrix can be diagonalized by a unitary rotation in the $1$--$3$ plane,
\begin{equation}
R =
\begin{pmatrix}
\cos\theta & \sin\theta\,e^{-i\psi}\\
-\sin\theta\,e^{i\psi} & \cos\theta
\end{pmatrix},
\label{eq:R13}
\end{equation}
where $\theta$ is the rotation angle and $\psi$ is a CP-violating phase.

The condition that the off-diagonal element of $R^TMR$ vanishes leads to
\begin{equation}
Y\cos2\theta
+\sin\theta\cos\theta\Big[
\cos\psi\,(X-Z)
-i\sin\psi\,(X+Z)
\Big]=0.
\end{equation}
This yields
\begin{equation}
\boxed{
\frac{\tan2\theta}{2}
=
\frac{Y}
{\cos\psi\,(Z-X)+i\sin\psi\,(X+Z)} }.
\label{eq:tan2theta_app}
\end{equation}

\subsection*{Dimensionless Parametrization}

To simplify the analysis, we normalize all quantities with respect to $b$ and
define
\begin{equation}
\frac{a}{b}=\lambda_2 e^{i\phi_{ab}},
\qquad
\frac{d}{b}=\lambda_1 e^{i\phi_{db}},
\end{equation}
with $b$ chosen real and positive. The dimensionless parameters then become
\begin{align}
\tilde X &= \lambda_2 e^{i\phi_{ab}}+1-\frac{\lambda_1}{2}e^{i\phi_{db}},\\
\tilde Y &= \frac{\sqrt{3}}{2}\lambda_1 e^{i\phi_{db}},\\
\tilde Z &= \lambda_2 e^{i\phi_{ab}}-1+\frac{\lambda_1}{2}e^{i\phi_{db}}.
\end{align}
From these expressions,
\begin{equation}
\tilde X+\tilde Z = 2\lambda_2 e^{i\phi_{ab}},
\qquad
\tilde Z-\tilde X = \lambda_1 e^{i\phi_{db}}-2.
\end{equation}

Substituting into Eq.~\eqref{eq:tan2theta_app} yields
\begin{equation}
\tan2\theta
=
\frac{\sqrt{3}\lambda_1 e^{i\phi_{db}}}
{2i\lambda_2 e^{i\phi_{ab}}\sin\psi
+(\lambda_1 e^{i\phi_{db}}-2)\cos\psi}.
\end{equation}
Separating the real part gives
\begin{equation}
\tan2\theta
=
\frac{\sqrt{3}\lambda_1\cos\phi_{db}}
{\lambda_1\cos\psi\cos\phi_{db}
-2(\lambda_2\sin\psi\sin\phi_{ab}+\cos\psi)} 
\label{eq:tan2theta_final}
\end{equation}

\subsection*{Determination of the CP Phase $\psi$}

Requiring the imaginary part of the denominator to vanish leads to
\begin{equation}
\lambda_2\sin\psi\cos(\phi_{ab}-\phi_{db})
+\cos\psi\sin\phi_{db}=0,
\end{equation}
which yields
\begin{equation}
\tan\psi
=
-\frac{\sin\phi_{db}}
{\lambda_2\cos(\phi_{ab}-\phi_{db})} \,.
\label{eq:tanpsi_final}
\end{equation}

\subsection*{Connection to PMNS Parameters}
Since the charged-lepton mass matrix is diagonal, $U_\ell=\mathbf{I}$, the PMNS
matrix coincides with the neutrino diagonalization matrix,
\begin{equation}
U_{\rm PMNS}=U_{\rm TBM}U_{13}(\theta,\psi)U_m.
\end{equation}
Comparing with the standard PDG parametrization yields the simplified relations can be established between neutrino mixing angles $\theta_{12}, \theta_{23}, \theta_{13}$ and the Dirac CP violating phase $\delta$ in terms of the model parameters $\theta, \psi$. The 
\begin{equation}
\sin\theta_{13}=\sqrt{\frac{2}{3}}\,|\sin\theta| 
\end{equation}
For $\sin\theta > 0$, one gets $\delta_{\rm CP}=\psi$
while for $\sin\theta<0$, $\delta_{\rm CP}=\psi+\pi$. In both cases,
$\tan\delta_{\rm CP}=\tan\psi$, and hence
\begin{equation}
\tan\delta_{\rm CP}
=
-\frac{\sin\phi_{db}}
{\lambda_2\cos(\phi_{ab}-\phi_{db})}\,.
\end{equation}

The remaining mixing angles follow as
\begin{eqnarray}
\sin\theta_{12} = 1/\big(\sqrt{3}\cos\theta_{13}\big)
\,, \quad 
\tan\theta_{23}
=
\frac{e^{-i\psi}\sin\theta+\sqrt{3}\,\cos\theta}
     {e^{-i\psi}\sin\theta-\sqrt{3}\,\cos\theta}.
\label{eqn:theta23}
\end{eqnarray}

These expressions demonstrate explicitly that all lepton mixing angles and the
Dirac CP phase are controlled by a single rotation angle $\theta$ and a single
phase $\psi$, which originate from the $A_4$-breaking structure of the sterile
sector in the double seesaw mechanism.

\bibliographystyle{JHEP_improved}
\bibliography{refsA4DSS}
\end{document}